# Solution-processed hybrid graphene flake/2H-MoS$_2$ quantum dot heterostructures for efficient electrochemical hydrogen evolution

Leyla Najafi$^{a,‡}$, Sebastiano Bellani$^{a,‡}$, Beatriz Martín-García$^{a,b}$, Reinier Oropesa-Nuñez$^{a}$, Antonio Esau Del Rio Castillo$^{a}$, Mirko Prato$^{c}$, Iwan Moreels$^{a,b}$, and Francesco Bonaccorso$^{*,a}$

$^a$ Graphene Labs, Istituto Italiano di Tecnologia, via Morego 30, 16163 Genova, Italy.

$^b$ Nanochemistry, Istituto Italiano di Tecnologia, via Morego 30, 16163 Genova, Italy.

$^c$ Materials Characterization Facility, Istituto Italiano di Tecnologia, via Morego 30, 16163 Genova, Italy.

* Corresponding authors: francesco.bonaccorso@iit.it.

$^‡$ These authors contributed equally

## Abstract

We designed solution-processed, flexible hybrid graphene flake/2H-MoS$_2$ quantum dot (QD) heterostructures, showing enhanced electrocatalytic activity for the hydrogen evolution reaction (HER) with respect to their native individual components. The 2H-MoS$_2$ QDs are produced through a scalable, environmentally friendly, one-step solvothermal approach from two-dimensional (2D) 2H-MoS$_2$ flakes obtained by liquid phase exfoliation (LPE) of their bulk counterpart in 2-Propanol. This QDs synthesis avoids the use of high boiling point and/or toxic solvents. Graphene flakes are produced by LPE of graphite in N-Methyl-2-pyrrolidone. The electrochemical properties of 2H-MoS$_2$ QDs and their HER-favorable chemical and electronic coupling with graphene enable to reach current density of 10 mA/cm² at an overpotential of 136 mV, surpassing the performances of graphene flake/2H-MoS$_2$ (1T-MoS$_2$) flake heterostructures. Our approach provides a shortcut, viable and cost-effective method for enhancing the 2D materials electrocatalytic properties.

## Introduction

Quantum dots (QDs) derived from the atomically-thin two-dimensional (2D) materials (2D-QDs), such as graphene, transition metal dichalcolgenides (TMDs), graphitic carbon nitride (g-C$_3$N$_4$), hexagonal boron nitride (h-BN) and monoatomic buckled crystals (e.g., phosphorene) are emerging as a new class of zero-dimensional (0D) materials.[1-4] In particular, their unique physical, (opto)electronic and electrochemical properties are promising for a wide range of applications, including optical imaging, energy conversion (e.g., (photo)electrocatalysis and photovoltaics), and energy storage (e.g., supercapacitors and batteries).[1,-4] Amongst the large variety of 2D-QDs, MoS$_2$

QDs have drawn particular attention because of their appealing photoluminescence (PL) emission,[5-8] *i.e.*, enabling the realization of biosensors[5-8] and electrocatalysts. The latter is particularly relevant for the hydrogen evolution reaction (HER) process.[9,10] Theoretical[11] and experimental studies[12,13] have identified that the unsaturated S atoms located at the edges of thermodynamically favourable phases of $MoS_2$, *i.e.*, stable, semiconducting 2H (trigonal prismatic), can absorb $H^+$ with a small Gibbs free energy ($\Delta G_H^0 \approx 0.08$ eV),[11] acting as active site for HER.[12] In addition, the electronic structure of the $MoS_2$ edge is dominated by metallic one-dimensional states,[14] differently from the semiconducting states in bulk $MoS_2$ and in basal planes of $MoS_2$ flakes,[12-14] where electrons transport is limited by hopping transport mechanisms.[14,15] Recent advances have also shown that the HER activity of $MoS_2$ flakes can be significantly enhanced when the semiconducting 2H phase of bulk $MoS_2$ is converted into metallic 1T (octahedral) phase via chemical exfoliation using lithium (Li)[16] or organo-lithium compounds.[17] However, the 1T-$MoS_2$ phase is thermodynamically metastable with a relaxation energy of ~1.0 eV[18] for the conversion to the stable 2H phase.[19] This limits the exploitation of 1T-$MoS_2$ for HER applications. Thus, extensive efforts are currently devoted toward the development of 2H-$MoS_2$ nanostructures to maximize the number of active edge sites[12,20] and/or to tailor the edge structure itself to reach enhanced HER kinetics.[12,21,22] Amongst different strategies proposed to advance the HER kinetics,[12,20,21,22] the exploitation of graphene-based scaffolds have been recently demonstrated to be a promising route to improve the electronic conductivity of 2H-$MoS_2$ flakes.[23] In fact, the use of graphene-based scaffolds afford to disperse 2H-$MoS_2$ flakes with a large amount of accessible active edges[24,25] enhancing the negative charge density, *i.e.,* to decrease their $\Delta G_H^0$ facilitating the HER process.[24] Moreover, graphene-based scaffolds represent free-standing flexible supports,[26] meeting the mechanical requirements for innovative energy conversion technologies, and, at the same time, offering interconnected porosities for the transport of evolved $H_2$.[27] In this context, flexible graphene flake/2H-$MoS_2$ QD heterostructures should offer a two-fold benefit.[24] On one hand, the QDs provide higher number of catalytic and conductive edge sites with respect to their native 2H-$MoS_2$ flakes.[9-10] On the other hand, flexible graphene-based conducting scaffold enables the electrons to access the active surface of QDs, thus favouring the HER kinetics.[24] Following this rationale, we design solution-processed flexible hybrid graphene flake/2H-$MoS_2$ QD heterostructures, showing enhanced electrocatalytic activity for HER with respect to their native individual components. So far, 2H-$MoS_2$ QDs have been synthetized by various methods, including ultrasonication,[8] chemical intercalation,[7,9] electrochemical exfoliation,[6] solvothermal cutting[5] and mechanical grinding.[10] However, the aforementioned methods suffer some limitations such as large size distribution[9] and multiple step processesing,[9,10] which are often carried out in high boiling point/toxic solvents (*e.g.*, dimethylformamide and N-Methyl-2-pyrrolidone (NMP)).[7-10]

Here, we produce $MoS_2$ QDs through a facile and scalable one-step solvothermal approach starting from 2H-$MoS_2$ flakes previously obtained by liquid phase exfoliation (LPE)[28,29] of bulk $MoS_2$ in 2-

Propanol (IPA). This material synthesis avoids the use of high boiling point and/or toxic solvents. Actually, the LPE is a promising approach for high-throughput industrial processes due to its low cost, simplicity and scalability.[30] Thus, by using the aforementioned methods and exploiting sedimentation-based separation (SBS)[31,32] we obtained sample I (2H-MoS$_2$ flakes) and sample II (MoS$_2$-QDs) (Scheme 1, see Supporting Information (S.I.) for the experimental details). The 1T-MoS$_2$ flakes are also produced by chemical lithium intercalation,[16] as HER catalyst benchmark,[16,17] see S.I. for details about production and characterization.

**Scheme 1.** Solvothermal synthesis of MoS$_2$ QDs in IPA.

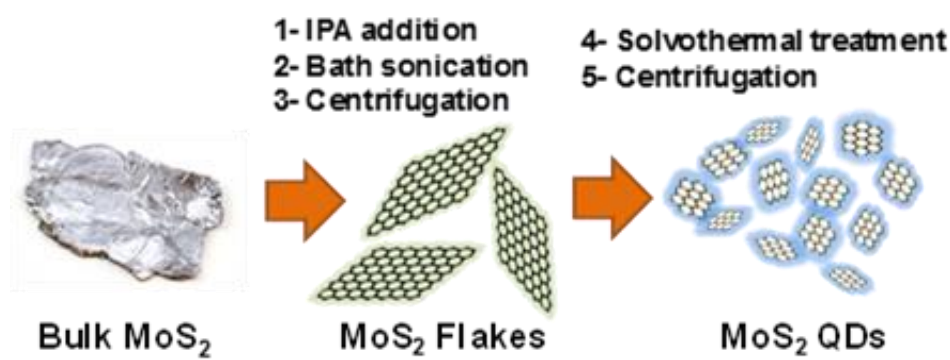

The lateral size and thickness of sample I and sample II are characterized by means of transmission electron microscopy (TEM) and atomic force microscopy (AFM). Figure 1a,b show representative TEM images of sample I and II, respectively. Sample I is composed by irregularly shaped flakes with average lateral dimension of ~420 nm (Figure 1c), while sample II consists of QDs with average lateral dimension of ~6 nm (Figure 1d). Figure 1e,f show representative AFM images of sample I and II, respectively. Height profiles (dashed white lines) indicate the presence of one- and two-layer flakes and QDs (the monolayer thickness is between 0.7-0.8 nm[33]). The average thickness of both the flakes and QDs is ~2.7 nm (Figure 1g,h).

X-ray photoelectron spectroscopy (XPS) measurements are carried out on both the as-produced flakes and QDs to determine their elemental composition and chemical phase. The S 2s and Mo 3d XPS spectra are shown in Figure 2a together with their deconvolution.[34] Here, the peak at the lowest binding energy (~226 eV) is assigned to S 2s while the peak at ~229eV is assigned to Mo 3d$_{5/2}$ of 2H-MoS$_2$.[34] The peak centered at ~232.5 eV can be fitted with two components.[34] The first component (~232eV) is assigned to Mo 3d$_{3/2}$ of the 2H-MoS$_2$.[34] Instead, the second component (~233eV), as well as the low intensity peak centered at ~236eV, are associated to the MoO$_3$ phase,[34,35] usually produced as a by-product of exfoliated MoS$_2$ flakes exposed to air.[35] However, the MoO$_3$-related peaks in both flakes and QDs spectra are negligible, indicating a transformation of only a small fraction of sulphide to oxide during the production of flakes and QDs from the bulk MoS$_2$. In fact, the percentage content (%c) of MoO$_3$ is <7% and <5% for flakes and QDs, respectively. These results prove that our production method produces 2H-MoS$_2$ flakes and QDs, overcoming the drawbacks of previous studies on MoS$_2$ flakes produced by LPE in NMP, where oxidized species (%c between 40-60%, depending on processing) are present.[34,35] Moreover, the solvothermal treatment does not change the chemical composition of the 2H-MoS$_2$ flakes, since there are no significant differences between the XPS spectra of flakes and QDs.

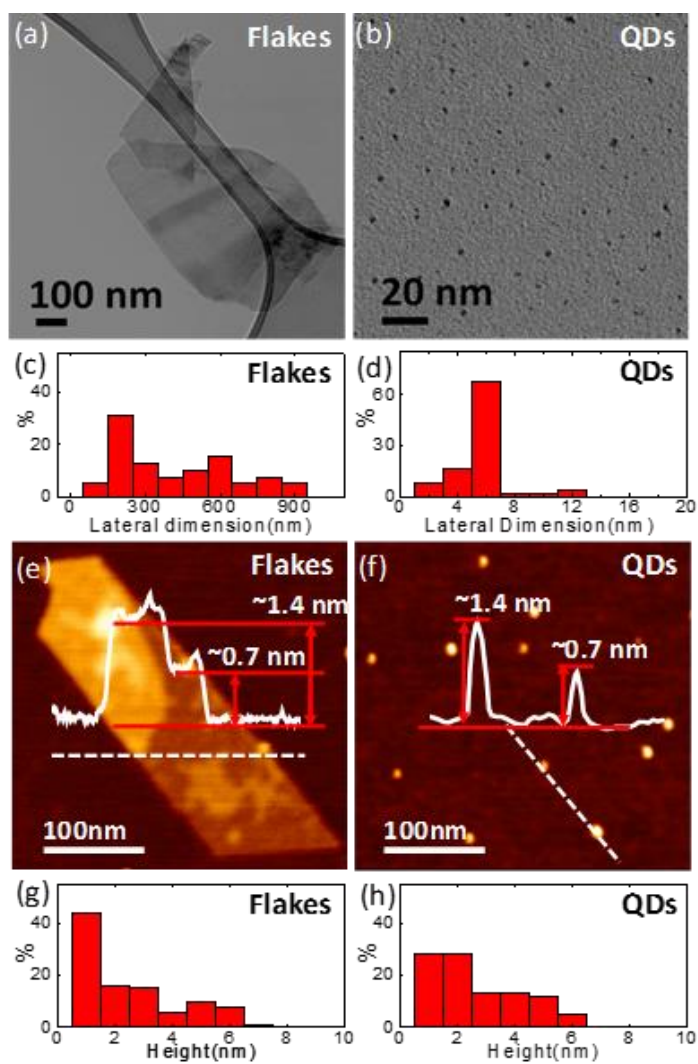

**Figure 1.** (a) TEM images of the as-produced flakes and (b) QDs and their corresponding statistical analysis of (c) flakes and (d) QDs. AFM images of representative as-produced (e) MoS$_2$ flakes and (f) QDs, deposited onto mica sheets. Representative height profiles (solid white lines) of the indicated sections (white dashed lines) are also shown. AFM statistical analysis of (g) flakes and (h) QDs.

X-ray diffraction (XRD) measurements are used to evaluate the crystal structure of 2H-MoS$_2$ flakes and QDs with respect to the bulk MoS$_2$ (Figure 2b). While bulk MoS$_2$ exhibits the characteristic XRD peaks of a hexagonal-structure polycrystalline films,[36] 2H-MoS$_2$ flakes and QDs only show the dominant (002) peak, centred at 14.4°, corresponding to the interlayer d-spacing of 0.614 nm.[10] This indicates that few-layer 2H-MoS$_2$ flakes and QDs have the same single crystal structure.[36] Additional details concerning the XRD analysis are reported in S.I.

The as-produced 2H-MoS$_2$ flakes and QDs are then studied by optical absorption spectroscopy (OAS) (Figure 2c) and photoluminescence (PL) emission measurements (Figure S1). The analysis of the corresponding data is reported in details in S.I. Briefly, the 2H-MoS$_2$ QDs do not show the typical excitonic peaks (*A, B, C, D*) of the 2H-MoS$_2$ flakes, with their absorption edge shifted towards lower wavelength with respect to latter (Figure 2c). This is a consequence of quantum confinement.[10] Furthermore, the 2H-MoS$_2$ QDs show excitation-dependent PL emission (Figure S1), which is due to the combination of quantum confinement effect[7] and edge state emission.[6,37]

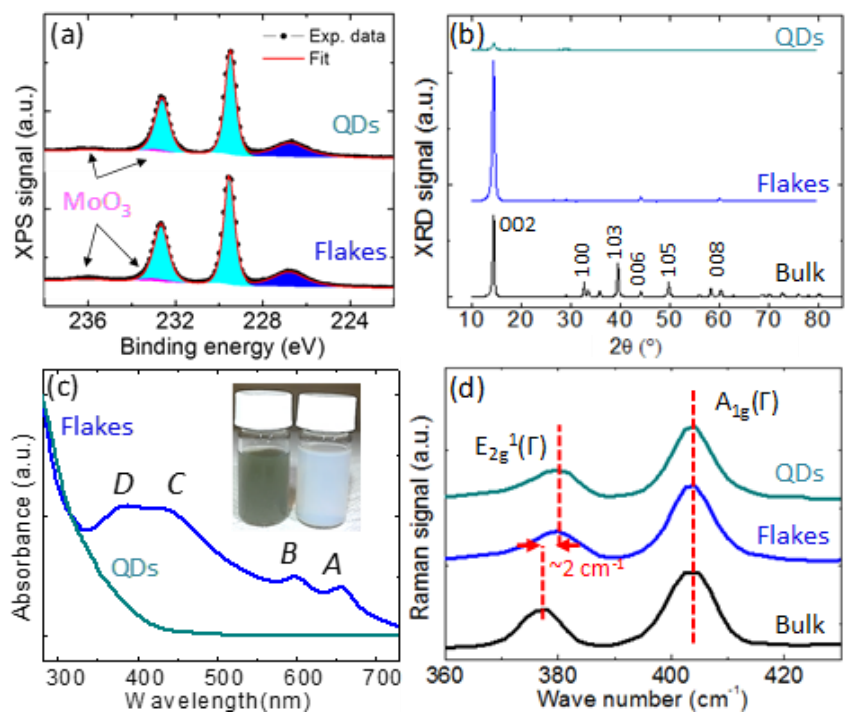

**Figure 2.** (a) Mo 3d and S 2s XPS spectra for 2H-MoS$_2$ flakes and QDs. Their deconvolution is also shown, and the area of the MoS$_2$ related bands are coloured in blue (S 2s) and cyan (Mo 3d). The bands attributed to the MoO$_3$ are coloured in magenta. (b) XRD spectra of bulk MoS$_2$, 2H-MoS$_2$ flakes and 2H-MoS$_2$ QDs. (c) Absorption spectra of 2H-MoS$_2$ flakes and 2H-MoS$_2$ QDs. Inset: photograph of 2H-MoS$_2$ flakes (left) and 2H-MoS$_2$ QDs (right) dispersions. (d) Raman spectra of the bulk MoS$_2$, 2H-MoS$_2$flakes and 2H-MoS$_2$ QDs.

Raman spectroscopy is carried out to investigate the vibrational modes of 2H-MoS$_2$ flakes and QDs with respect to those of bulk MoS$_2$ (Figure 2d). Representative Raman spectra show the presence of first-order modes at the Brillouin zone center $E_{2g}^1(\Gamma)$ (~379 cm$^{-1}$ for both 2H-MoS$_2$ flakes and QDs, and ~377 cm$^{-1}$ for bulk MoS$_2$) and $A_{1g}(\Gamma)$ (~403 cm$^{-1}$) involving the in-plane displacement of Mo and S atoms and the out-of-plane displacement of S atoms, respectively.[38,39] The $E_{2g}^1(\Gamma)$ mode of both the 2H-MoS$_2$ flakes and QDs exhibits softening with respect to the one of the bulk MoS$_2$. The shift of the $E_{2g}^1(\Gamma)$ mode is explained by the dielectric screening of long range Coulomb MoS$_2$ interlayer interaction.[38] Additional discussion is reported in S.I., together with the statistical analysis of the data (Figure S2). Extended Raman spectra between 200-900 cm$^{-1}$ (Figure S3) do not reveal additional peaks related to molybdenum oxide species,[40] in agreement with XPS (Figure 2a) and XRD (Figure 2b) data. We then designed flexible hybrid heterostructures of graphene flakes and $MoS_2$ flakes or quantum dots (named as graphene/2H-$MoS_2$ flakes and graphene/2H-$MoS_2$ QDs, respectively). Single- and few-layer graphene flakes are produced by LPE of pristine graphite in NMP[28,31] (see S.I. for experimental and characterization details, Figure S4-s8). The graphene flake/2H-$MoS_2$ flake (or QDs) heterostructures are obtained by depositing sequentially graphene flakes and $MoS_2$ flakes (or QDs) dispersions on nylon membranes through vacuum filtration (see S.I., Figure S9). Notably, this solution-processed fabrication of heterostructures is scalable and compatible with high-throughput industrial processes.

The HER electrocatalytic activity of the heterostructures is investigated in 0.5 M $H_2SO_4$. The 2H-$MoS_2$ flakes and QDs are also deposited and tested on a glassy carbon (GC) electrode (*i.e.*, GC/2H-$MoS_2$ flakes and GC/2H-$MoS_2$ QDs) in order to provide their individual electrocatalytic properties on a flat inert conductive substrate. Moreover, GC/1T-$MoS_2$ flakes and graphene/1T-$MoS_2$ flakes are fabricated and tested as benchmark for HER[16,17] (see S.I. for experimental and characterization details, Figure S10). It is worth to note that the synthesis of QDs starting from 1T-$MoS_2$ flakes leads to 2H-$MoS_2$ QDs, as indicated by XPS analysis (Figure S10). This can be ascribed to the intrinsic metastable nature of 1T-$MoS_2$ flakes,[18] which relax, during the solvothermal treatment, towards the thermodynamically favored 2H phase.[18,19] Figure 3a displays the *iR*-corrected polarization curves for the different $MoS_2$-based electrodes. The commonly used figure of merit (FoM) to evaluate the HER performance of an electrocatalyst is the overpotential at 10 mA/cm$^2$ cathodic current density ($\eta_{10}$).[41] The GC/2H-$MoS_2$ QDs electrode shows ~60 mV lower $\eta_{10}$ (~312 mV) with respect to that of the GC/2H-$MoS_2$ flakes (~372 mV). The HER activity increases remarkably in the case of the heterostructures ($\eta_{10}$ of ~175 mV and ~136 mV for the graphene/2H-$MoS_2$ flakes and graphene/2H-$MoS_2$ QDs, respectively), with respect to that obtained on GC. Moreover, the $\eta_{10}$ of graphene/2H-$MoS_2$ QDs is lower with respect to those of GC/1T-$MoS_2$ flakes (~235 mV) and graphene/1T-$MoS_2$ flakes (~151 mV). The Tafel slope and the exchange current density ($j_0$) are also useful FoM to assess the performance of catalysts for HER.[41] Tafel slope is used to evaluate the reaction processes of HER,[41] while $j_0$ is positively correlated to the number of the catalytic active sites and their HER kinetics.[41] The obtained Tafel slopes are ∼145, ∼98, and ∼78 mV/dec for GC/2H-$MoS_2$ flakes, GC/2H-$MoS_2$ QDs and GC/1T-$MoS_2$ flakes, respectively. These values suggest that the HER is limited by the Volmer reaction ($H_3O^+ + e^- \rightarrow H_{ad} + H_2O$, where $H_{ad}$ refer to adsorbed hydrogen atoms and the theoretical Tafel slope is 120 mV/dec)[41] for GC/2H-$MoS_2$ flakes. In fact, for this configuration, the limited number of edges slows down the hydrogen adsorption rate.[41] Differently, for the case of GC/1T-$MoS_2$ flakes, the Volmer-Heyrovsky mechanism (*i.e.,* Volmer and Heyrovsky reactions: $H_{ad} + H_3O^+ + e^- \rightarrow H_2 + H_2O$, where Tafel slope is between 120 and 40 mV/dec)[41] occurs for HER. The observed behaviour is in agreement with previous literature on 2H-$MoS_2$ QDs,[9,10] where the reaction kinetics are not limited by the number of catalytic edge sites as for the 2H-MoS$_2$ flakes and, therefore, the Heyrovsky-Volmer mechanism is thus facilitated.[10] The Tafel slopes for graphene/2H-MoS$_2$ flakes, graphene/2H-MoS$_2$ QDs and graphene/1T-MoS$_2$ flakes are ∼163, ∼141, and ∼82 mV/dec, respectively. These values increase of ~0.13%, 0.44% and 0.05% with respect to those for GC/2H-MoS$_2$ flakes, GC/2H-MoS$_2$ QDs and GC/1T-MoS$_2$ flakes, respectively. This could be ascribed to the presence of large H$_{ad}$ coverage for the range of η where the Tafel slope has been extrapolated,[42] in agreement with theoretical simulation of HER kinetics in acidic conditions.[42] The current densities (between 1.5 and 6.5mA/cm$^2$) measured in the 0.10-0.05 V *vs.* RHE range suggest the presence of an intermediate step, ascribable to the Volmer reaction, preceding the H$_2$ evolution. Moreover, the j$_0$ values for graphene/2H-MoS$_2$ flakes and graphene/2H-MoS$_2$ QDs (~0.97 mA/cm$^2$ and ~1.31 mA/cm$^2$, respectively) are significantly increased with respect to the one obtained for GC/2H-MoS$_2$ flakes and GC/2H-MoS$_2$ QDs (~0.03 mA/cm$^2$). The electrocatalytic stability of the graphene/2H-MoS$_2$ QDs in HER-condition is evaluated by chronoamperometry measurements (j-t curves) at -0.5 V *vs.* RHE (Figure S11). The graphene/2H-MoS$_2$ QDs show a progressive HER activation, with a ~10% increase of j after 200 min, while the current density of graphene/2H-MoS$_2$ flakes decreases of ~4% with respect to the starting values.

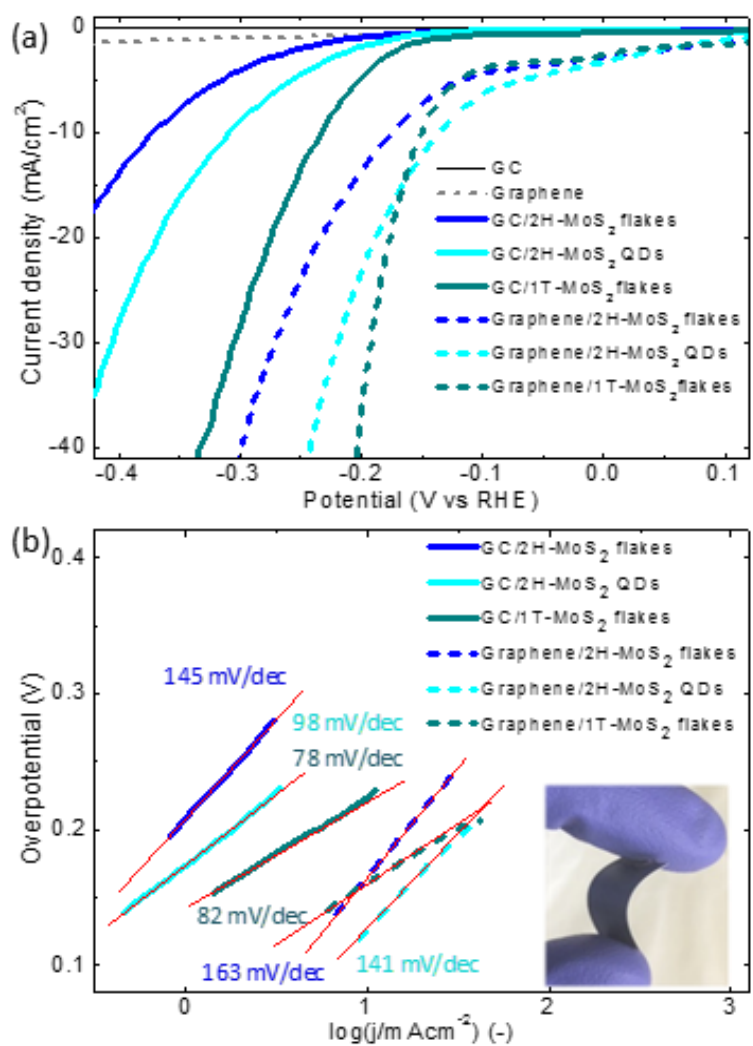

**Figure 3.** (a) Polarization curves of 2H-MoS$_2$ flakes, 2H-MoS$_2$ QDs, 1T-MoS$_2$ flakes, on GC electrode (solid lines), and graphene/2H-MoS$_2$ flakes, graphene/2H-MoS$_2$ QDs and graphene/1T-MoS$_2$ flakes (dashed lines). Polarization curves of GC and graphene flakes are also shown for comparison. (b) Tafel plots of the same MoS$_2$-based electrodes shown in panel (a). Linear fits (red lines) and the corresponding Tafel slope values are reported. Inset: photograph of a representative flexible hybrid graphene flakes/2H-MoS$_2$ QDs electrode.

These results suggest that the catalytic edge sites of 2H-MoS$_2$ QDs are more resistant toward oxidative/degradation processes, which passivate the HER catalytic sites, compared to those of 2H-MoS$_2$.[10]

In conclusion, we have shown that solution-processed flexible graphene flake/2H-MoS$_2$ QD heterostructures yield remarkable and stable HER electrocatalytic activity, overwhelming the one of GC/2H-MoS$_2$ flakes and GC/2H-MoS$_2$ QDs, respectively. The 2H-MoS$_2$ QDs, produced by an environmentally friendly solvothermal process in IPA, show average diameter of 6 nm, crystallinity retention, and low percentage content (<5%) of oxidized by-products. The hybrid graphene flakes/2H-MoS$_2$ QDs heterostructure enables to reach a lower η$_{10}$ (~136 mV) with respect to that of both graphene flakes/2H-MoS$_2$ flakes (175 mV) and GC/1T-MoS$_2$ flakes heterostructures (~235 mV) as well as graphene flakes/1T-MoS$_2$ flakes (~151 mV) heterostructure. These HER electrocatalytic performances approach those of several MoS$_2$-based catalyst reported in literature, overcoming those of recent MoS$_2$ flakes or MoS$_2$ QDs synthetized by scalable routes[10] (see S.I., Table S1). The as-produced 2H-MoS$_2$ QDs and their flexible heterostructures with graphene flakes are promising for viable and cost-effective HER-catalysts. Our results can boost the development of other hybrid heterostructures between 2D materials and 2D-QDs, providing a facile scalable method for enhancing the electrocatalytic properties of energy conversion technologies.

## Associated contents

**Supporting Information**. Experimental methods; additional OAS, Raman, TEM, AFM, XPS and electrochemical characterizations of MoS$_2$ (bulk, flakes and QDs), graphene flakes and hybrid electrodes. This material is available free of charge via the Internet at http://pubs.acs.org.

**Funding Sources**

This project has received funding from the European Union's Horizon 2020 research and innovation program under grant agreement No. 696656-GrapheneCore1.

## References
(1) Wang, X.; Sun, G.; Li, N; Chen, P. Quantum dots derived from two-dimensional materials and their applications for catalysis and energy. *Chem. Soc. Rev.* **2016**, *45*, 2239-2262.

(2) Ferrari A. C. et al., Science and technology roadmap for graphene, related two-dimensional crystals, and hybrid systems. *Nanoscale*, **2015**, *7*, 4598-4810.

(3) Zhang, Z.; Zhang, J.; Chen, N.; Qu, L. Graphene quantum dots: an emerging material for energy-related applications and beyond. *Energy Environ. Sci.* **2012**, *5*, 8869-8890.

(4) Bonaccorso, F.; Colombo, L.; Yu, G.; Stoller, M.; Tozzini, V.; Ferrari, A. C.; Ruoff, R. S.; Pellegrini, V. Graphene, related two-dimensional crystals, and hybrid systems for energy conversion and storage. *Science*, **2015**, *347*, 1246501.

(5) Xu, S.; Li, D.; Wu, P. One-pot, facile, and versatile synthesis of monolayer MoS2/WS2 quantum dots as bioimaging probes and efficient electrocatalysts for hydrogen evolution reaction. *Adv. Funct. Mater.* **2015**, *25*, 1127.

(6) Gopalakrishnan, D.; Damien, D.; Li, B.; Gullappalli, H.; Pillai, V. K.; Ajayan, P. M.; Shaijumon, M. M. Electrochemical synthesis of luminescent MoS2 quantum dots. *Chem. Commun.* **2015**, *51*, 6293-6296.

(7) Wu, J. Y.; Zhang, X. Y.; Ma, X. D.; Qiu, Y. P.; Zhang, T. High quantum-yield luminescent MoS2 quantum dots with variable light emission created via direct ultrasonic exfoliation of MoS2 nanosheets. *RSC Adv.* **2015**, *5*, 95178-95182.

(8) Yue, N.; Weicheng, J.; Rongguo, W.; Guomin, D.; Yifan, H. Hybrid nanostructures combining graphene–MoS2 quantum dots for gas sensing. *J. Mater. Chem. A*, **2016**, *4*, 8198.

(9) Qiao, W.; Yan, S.; Song, X.; Zhang, X.; Sun, Y.; Chen, X.; Zhong, W.; Du, Y. Monolayer MoS2 quantum dots as catalysts for efficient hydrogen evolution. *RSC Adv.*, **2015**, *5*, 97696-97701

(10) Benson, J.; Li, M.; Wang, S.; Wang, P.; Papakonstantinou, P. Electrocatalytic hydrogen evolution reaction on edges of a few layer molybdenum disulfide nanodots, *ACS Appl. Mater. Interfaces*, **2015**, 7,14113-14122.

(11) Jaramillo, T. F.; Jørgensen, K. P.; Bonde, J.; Nielsen, J. H.; Horch, S.; Chorkendorff, I. Identification of active edge sites for electrochemical H2 evolution from MoS2 nanocatalysts. *Science*, **2007**, *317*, 100-102.

(12) Kibsgaard, J.; Chen, Z.; Reinecke, B. N.; Jaramillo, T. F. Engineering the surface structure of MoS2 to preferentially expose active edge sites for electrocatalysis. *Nat. Mater.* **2012**, *11*, 963-969.

(13) Laursen, A. B.; Kegnæs, S.; Dahl, S.; Chorkendorff, I. *Energy Environ. Sci.* **2012**, *5*, 5577-5591.ì

(14) Bollinger, M. V.; Lauritsen, J. V.; Jacobsen, K. W.; Nørskov, J. K.; Helveg, S.; Besenbacher, F. One-dimensional metallic edge states in MoS2. *Phys. Rev. Lett*. **2001**, *87*, 196803.

(15) Yu, Y.; Huang, S.; Li, Y.; Steinmann, S.; Yang, W.; Cao, L.; Layer-dependent electrocatalysis of MoS2 for hydrogen evolution. *Nano Lett*. **2014**, *14*, 553-558.

(16) Voiry, D.; Salehi, M.; Silva, R.; Fujita, T.; Chen, M.; Asefa, T., Shenoy, V.B.; Eda, G. T.; Chhowalla, M.; Conducting MoS2 nanosheets as catalysts for hydrogen evolution reaction. Nano Lett. **2013**, *13*, 6222-6227.

(17) Chhowalla, M.; Shin, H. S.; Eda, G.; Li, L. J.; Loh, K. P.; Zhang, H.; The chemistry of two-dimensional layered transition metal dichalcogenide nanosheets. *Nature Chem.*, **2013**, *5*, 263.

(18) Gao, G.; Jiao, Y.; Ma, F.; Jiao, Y.; Waclawik, E.; Du, A. Charge mediated semiconducting-to-metallic phase transition in molybdenum disulfide monolayer and hydrogen evolution reaction in new 1T' phase. *J. Phys. Chem. C*, **2015**, *119*, 13124-13128.

(19) Tsai, H. L.; Heising, J.; Schindler, J. L.; Kannewurf, C. R.; Kanatzidis, M. G. Exfoliated-restacked phase of WS2. *Chem. Mater.*, **1997**, *9*, 879-882.

(20) Chung, D. Y.; Park, S. K.; Chung, Y. H.; Yu, S. H.; Lim, D. H.; Jung, N.; Ham, H. C.; Park, H. Y.; Piao, Y.; Yoo, S. J.; Sung, Y. E. Edge-exposed MoS2 nano-assembled structures as efficient electrocatalysts for hydrogen evolution reaction. *Nanoscale*, **2014**, *6*, 2131-2136.

(21) Hu, J.; Huang, B.; Zhang, C.; Wang, Z.; An, Y.; Zhou, D.; Lin, H.; Leung, M. K.; Yang, S. Engineering stepped edge surface structures of MoS2 sheet stacks to accelerate the hydrogen evolution reaction. *Energy Environ. Sci*. **2017**, *10*, 593-603.

(22) Zhang, J.; Wang, T.; Liu, P.; Liu, S.; Dong, R.; Zhuang, X.; Chen, M.; Feng, X. Engineering water dissociation sites in MoS2 nanosheets for accelerated electrocatalytic hydrogen production. *Energy Environ. Sci*. **2016**, 9, 2789-2793.

(23) Yang, J.; Shin, H.S., Recent advances in layered transition metal dichalcogenides for hydrogen evolution reaction. *J. Mater. Chem. A*, **2014**, 2, 5979-5985.

(24) Li, Y.; Wang, H.; Xie, L.; Liang, Y.; Hong, G.; Dai, H. MoS2 nanoparticles grown on graphene: an advanced catalyst for hydrogen evolution reaction. *J. Am. Chem. Soc*. **2011**, 133, 7296.

(25) Yang, J.; Voiry, D.; Ahn, S. J.; Kang, D.; Kim, A. Y.; Chhowalla, M.; Shin, H. S.; Two-Dimensional Hybrid Nanosheets of Tungsten Disulfide and Reduced Graphene Oxide as Catalysts for Enhanced Hydrogen Evolution. *Angew. Chem. Int. Ed*. **2013**, *52*, 13751-13754.

(26) Hou, Y.; Qiu, M.; Zhang, T.; Ma, J.; Liu, S.; Yuan, C; Feng, X. Efficient Electrochemical and Photoelectrochemical Water Splitting by a 3D Nanostructured Carbon Supported on Flexible Exfoliated Graphene Foil., *Adv. Mater.,* **2017**, *29*, 1604480.

(27) Hou, D.; Zhou, W.; Zhou, K.; Zhou, Y.; Zhong, J.; Yang, L.; Lu, J.; Li, G.; Chen, S. Flexible and porous catalyst electrodes constructed by Co nanoparticles@nitrogen-doped graphene films for highly efficient hydrogen evolution. *J. Mater. Chem. A*, **2015**, *3*, 15962-15968.

(28) Bonaccorso, F.; Lombardo, A.; Hasan, T.; Sun, Z.; Colombo, L.; Ferrari, A. C. Production and processing of graphene and 2d crystals. *Mater. Today*, **2012**, *15*, 564-589.

(29) Ciesielski, A.; Samorì, P. Graphene via sonication assisted liquid-phase exfoliation. *Chem. Soc. Rev*., **2014**, *43*, 381-398.

(30) Bonaccorso, F.; Bartolotta, A.; Coleman, J. N.; Backes, C. 2D-Crystal-Based Functional Inks. *Adv. Mater.*, **2016**, *28*, 6136.

(31) Maragó, O. M.; Bonaccorso, F.; Saija, R.; Privitera, G.; Gucciardi, P. G.; Iati, M. A.; Calogero, G.; Jones, P. H.; Borghese, F.; Denti, P.; Nicolosi, V. Brownian motion of graphene. *ACS Nano*, **2010**, *4*, 7515-7523

(32) Casaluci, S.; Gemmi, M.; Pellegrini, V.; Di Carlo, A.; Bonaccorso, F. Graphene-based large area dye-sensitized solar cell modules. N*anoscale*, **2016**, *8*, 5368.

(33) Halim, U.; Zheng, C. R.; Chen, Y.; Lin, Z.; Jiang, S.; Cheng, R.; Huang, Y.; Duan, X. A rational design of cosolvent exfoliation of layered materials by directly probing liquid–solid interaction. *Nat. Commun*. **2013**, *4*, 2213.

(34) Capasso, A.; Matteocci, F.; Najafi, L.; Prato, M.; Buha, J.; Cinà, L.; Pellegrini, V.; Carlo, A. D; Bonaccorso, F. Few-Layer MoS2 Flakes as Active Buffer Layer for Stable Perovskite Solar Cells. *Adv. Energy Mater*. **2016**, *6*, 1600920.

(35) Jawaid, A.; Nepal, D.; Park, K.; Jespersen, M.; Qualley, A.; Mirau, P.; Drummy, L. F.; Vaia, R. A. Mechanism for liquid phase exfoliation of MoS2. *Chem. Mater*. **2016**, *28*, 337-348.

(36) Joensen, P.; Crozier, E. D.; Alberding, N.; Frindt, R. F. A study of single-layer and restacked MoS2 by X-ray diffraction and X-ray absorption spectroscopy. *J. Phys. C: Solid State Phys*. **1987**, *20*, 4043.

(37) Gan, Z.; Xu, H.; Hao, Y. Mechanism for excitation-dependent photoluminescence from graphene quantum dots and other graphene oxide derivates: consensus, debates and challenges. *Nanoscale*, **2016**, *8*, 7794-7807.

(38) Saito, R.; Tatsumi, Y.; Huang, S.; Ling, X.; Dresselhaus, M. S., Raman spectroscopy of transition metal dichalcogenides. *J. Phys.: Condens. Matter*, **2016**, *28*, 353002.

(39) Bellani, S.; Najafi, L.; Capasso, A.; Castillo, A. E. D. R.; Antognazza, M. R.; Bonaccorso, F. Few-layer MoS2 flakes as a hole-selective layer for solution-processed hybrid organic hydrogen-evolving photocathodes. *J. Mater. Chem. A*, **2017**, *5*, 4384-4396.

(40) Dieterle, M.; Weinberg, G.; Mestl, G. Raman spectroscopy of molybdenum oxides Part I. Structural characterization of oxygen defects in MoO3–x by DR UV/VIS, Raman spectroscopy and X-ray diffraction. *Phys. Chem. Chem. Phys*., **2002**, *4*, 812-821.

(41) Vilekar, S. A.; Fishtik, I.; Datta, R. Kinetics of the hydrogen electrode reaction. *J. Electrochem. Soc*. **2010**, *157*, B1040-1050.

(42) Conway, B. E.; Gileadi, E. Kinetic theory of pseudo-capacitance and electrode reactions at appreciable surface coverage. *Trans. Faraday Soc*., **1962**, *58*, 2493-2509.

# Supporting Information

## S.1 Experimental Methods
### S.1.1 Synthesis
2H-MoS$_2$ quantum dots (QDs) are produced through an one-step solvothermal method starting from 2H-MoS$_2$ flakes, produced by liquid phase exfoliation (LPE)[1,2,3] of bulk MoS$_2$ crystals in 2-Propanol (IPA) followed by sedimentation-based separation (SBS).[Error! Bookmark not defined.,4,5] In detail, 30 mg of MoS$_2$ bulk crystal (Sigma Aldrich) are added to 50 mL of IPA and then ultrasonicated (Branson® 5800 cleaner, Branson Ultrasonics) for 8 h. The resulting dispersion is ultracentrifuged (Optima™ XE-90 ultracentrifuge, Beckman Coulter) for 15 min at 2700 $g$, in order to separate the un-exfoliated and thick MoS$_2$ crystals (collected as sediment) from the thinner 2H-MoS$_2$ flakes that remain in the supernatant. An aliquot (*i.e.* 10 mL) of the as-obtained 2H-MoS$_2$ flakes dispersion is kept for further characterization (sample I), while the rest is refluxed in air under stirring for 24 h at 140 °C. The resulting dispersion is subsequently ultracentrifuged for 30 min at 24600 $g$. Afterward, the supernatant is collected, thus obtaining the 2H-MoS$_2$ QDs dispersion (sample II).

The 1T-MoS$_2$ flakes are produced by a chemical lithium intercalation method.[Error! Bookmark not defined.,6] Briefly, 500 mg of MoS$_2$ bulk crystals are suspended in 10 mL of 1.6 M methyllithium (CH$_3$Li) in diethylether (Sigma Aldrich) and 10 mL of 1.6 M n-butyllithium (n-BuLi) in cyclohexane (Sigma Aldrich). The dispersion is stirred for 3 days at room temperature under argon atmosphere. The Li-intercalated material (Li$_x$MoS$_2$) ws separated by vacuum filtration under Ar atmosphere.[7] Li$_x$MoS$_2$ is washed in anhydrous hexane to remove non-intercalated Li ions and organic residues. Li$_x$MoS$_2$ powder is then exfoliated by ultrasonication in deionized (DI) water for 1 h. The dispersion is ultrasonicated for 10 min and then ultracentrifuged for 40 min at 67000 $g$. The collected precipitate is rinsed with MilliQ water and re-dispersed in 400 ml of IPA. The resulting dispersion is ultrasonicated for 30 min and ultracentrifuged for 20 min at 17000 $g$. The supernatant is then collected, thus obtaining the 1T-MoS$_2$ flake dispersion.

The graphene flakes dispersion is produced by LPE of graphite in N-Methyl-2-pyrrolidone (NMP).[Error! Bookmark not defined.-5] 1 g of graphite is dispersed in 100 ml of NMP and ultrasonicated for 6 hours. The obtained dispersion is ultracentrifuged at 17000 $g$ for 50 min at 15 °C. Finally, 80% of the supernatant is collected by pipetting, thus obtaining graphene flakes dispersion.

### S.1.2 Material characterization
Transmission electron microscopy (TEM) images are taken with a JEM 1011 (JEOL) transmission electron microscope, operating at 100 kV. Samples for the TEM measurements are prepared by drop-casting the 2H-MoS$_2$ flakes, 2H-MoS$_2$ QDs, 1T-MoS$_2$ flakes and graphene flake dispersions onto carbon-coated copper grids. Their lateral dimensions are measured using ImageJ software (NIH). Statistical TEM analysis is carried out by means of Origin 8.1 software (OriginLab).

Atomic force microscopy (AFM) images are taken using a Nanowizard III (JPK Instruments, Germany) mounted on an Axio Observer D1 (Carl Zeiss, Germany) inverted optical microscope. The AFM measurements are carried out by using PPP-NCHR cantilevers (Nanosensors, USA) with a nominal tip diameter of 10 nm. A drive frequency of ~295 kHz is used. Intermittent contact mode AFM images of 2.5×2.5 µm$^2$ and 500×500 nm$^2$ are collected with 512 data points per line and the working set point is kept above 70% of the free oscillation amplitude. The scan rate for acquisition of images is 0.7 Hz. Height profiles are processed by using the JPK Data Processing software (JPK Instruments, Germany) and the data are analyzed with OriginPro 9.1 software. Statistical AFM analysis is carried out by means of Origin 8.1 software (OriginLab) on four different AFM images for each sample. The samples are prepared by drop-casting 2H-MoS$_2$ flakes, 2H-MoS$_2$ QDs, 1T-MoS$_2$ flakes and graphene flake dispersions onto mica sheets (G250-1, Agar Scientific Ltd., Essex, U.K.).

X-ray photoelectron spectroscopy (XPS) is carried out on a Kratos Axis UltraDLD spectrometer, using a monochromatic Al K$_\alpha$ source (15 kV, 20 mA). The spectra are taken on a 300 µm x 700 µm area. Wide scans are collected with constant pass energy of 160 eV and energy step of 1 eV; high-resolution spectra are acquired at constant pass energy of 10 eV and energy step of 0.1 eV. The binding energy scale is internally referenced to the C 1s peak at 284.8 eV. The spectra are analyzed using the CasaXPS software (version 2.3.17). The fitting of the spectra is performed by using a linear background and Voigt profiles. The samples are prepared by drop-casting dispersions of 2H-MoS$_2$ flakes, 2H-MoS$_2$ QDs, 1T-MoS$_2$ flakes and graphene flakes onto Si/SiO$_2$ substrate (LDB Technologies Ltd). The graphene flakes sample is also annealed at 350 C° in order to remove residual NMP.

The crystal structure is characterized by X-ray diffraction (XRD) using a PANalytical Empyrean with CuKa radiation. The samples for XRD are prepared by drop-casting 2H-MoS$_2$ flakes, 2H-MoS$_2$ QDs, and 1T-MoS$_2$ flakes dispersions on a silicon wafer and dried under vacuum.

The optical absorption spectroscopy (OAS) measurements are carried out using a Cary Varian 6000i UVvis-NIR spectrometer using quartz cuvette with a path length of 1 cm. The 2H-MoS2 flakes, 2H-MoS$_2$ QDs and 1T-MoS$_2$ flakes are characterized as-produced, while for the graphene flakes, a 1:10 dilution of the corresponding dispersion is measured in order to avoid scattering losses. The corresponding solvent baselines are subtracted.

The steady-state photoluminescence (PL) emission measurements are performed using an Edinburgh Instruments FLS920 spectrofluorometer. The PL spectra are collected exciting the samples at different wavelengths ranging from 280 to 500 nm at a step of 20 nm, using a Xe lamp coupled to a monochromator. The 2H-MoS$_2$ QDs dispersions are contained in a quartz glass cuvette with a path length of 1 cm. To discard any contribution from the solvent (isopropanol), blank (control) measurement is carried out in the same experimental conditions used for the characterization of the aforementioned samples.

Raman measurements are carried out by using a Renishaw microRaman invia 1000 using a 50× objective, with an excitation wavelength of 532 nm and an incident power on the samples of 1 mW. For each sample, 50 spectra are collected. The sample are prepared by drop casting dispersions of 2H-MoS$_2$ flakes, 2H-MoS$_2$ QDs, and 1T-MoS$_2$ flakes and graphene flakes onto Si/SiO$_2$ (300 nm SiO$_2$) substrates and dried under vacuum. The spectra are fitted with Lorentzian functions. Statistical analysis is carried out by means of Origin 8.1 software (OriginLab).

### S.1.3 Fabrication of the electrodes

Dispersions of 2H-MoS$_2$ flakes, 2H-MoS$_2$ QDs, and 1T-MoS$_2$ flakes are deposited on glassy carbon (GC) sheets (Sigma Aldrich) (GC/2H-MoS$_2$ flakes, GC/2H-MoS$_2$ QDs and GC/1T-MoS$_2$ flakes, respectively) by drop-casting (mass loading of 0.5 mg/cm$^2$). Flexible hybrid heterostructures of graphene flakes (graphene), graphene flakes/2H-MoS$_2$ flakes (graphene/2H-MoS$_2$ flakes) and graphene flakes/2H-MoS$_2$ QDs (graphene/2H-MoS$_2$ QDs) or 1T-MoS$_2$ flakes (graphene/1T-MoS$_2$ flakes) are fabricated by sequentially depositing graphene flakes and MoS$_2$ flakes or QDs dispersions onto nylon membranes with size pore of 0.2 μm (Whatman® membrane filters nylon, Sigma Aldrich) through a vacuum filtration process (MoS$_2$ mass loading of 0.5 mg/cm$^2$). All the electrodes are dried overnight at room temperature before their electrochemical characterization.

### S.1.4 Electrodes characterization

The AFM images of the flexible electrodes (*i.e.,* graphene, graphene/2H-MoS$_2$ flakes and the graphene/2H MoS$_2$ QDs) are taken using the same set-up used for AFM characterization of materials.

Electrochemical measurements on the as-prepared electrodes are carried out at room temperature in a flat-bottom fused silica cell under a three-electrode configuration using CompactStat potentiostat/galvanostat station (Ivium), controlled via Ivium's own IviumSoft. A Pt wire is used as the counter-electrode and saturated KCl Ag/AgCl is used as the reference electrode. Measurements are carried out in 200 mL 0.5 M H$_2$SO$_4$ (99.999% purity, Sigma Aldrich) (pH 1). Oxygen is purged from electrolyte by flowing N$_2$ gas throughout the liquid volume using a porous frit for 30 minutes before starting the measurements. A constant, N$_2$ flow is maintained afterwards for the whole duration of the experiments, to avoid re-dissolution of molecular oxygen in the electrolyte. Potential difference between the working electrode and the Ag/AgCl reference electrode is converted to the reversible hydrogen electrode (RHE) scale via the Nernst equation: $E_{RHE} = E_{Ag/AgCl} + 0.059pH + E^0_{Ag/AgCl}$, where $E_{RHE}$ is the converted potential versus RHE, $E_{Ag/AgCl}$ is the experimental potential measured against the Ag/AgCl reference electrode, and $E^0_{Ag/AgCl}$ is the standard potential of Ag/AgCl at 25 °C (0.1976 V). Polarization curves are acquired at a 5 mV/s scan rate. Polarization curves from all catalysts are *iR*-corrected, where *i* is the current and the *R* is the series resistance arising from the substrate and electrolyte resistances. R is measured by electrochemical impedance spectroscopy (EIS) at open circuit potential and at frequency of $10^4$ Hz.

The linear portions of the Tafel plots are fit to the Tafel equation η= *b*\*log(j) + *A*,[8,9] where η is the overpotential with respect to the reversible hydrogen electrode potential (RHE), j is the current density, *b* is the Tafel slope and *A* is the intercept of the linear regression. The $j_0$ is the current calculated from the Tafel equation by setting η equal to zero. Stability tests are carried out by chronoamperometry measurements (*j-t* curves), *i.e.,* by measuring the current in potentiostatic mode at -0.5 V *vs.* RHE in 0.5 M $H_2SO_4$ over time (200 min).

### S.2 X-ray diffraction analysis of bulk $MoS_2$, 2H-$MoS_2$ flakes and 2H-$MoS_2$ QDs

X-ray diffraction (XRD) measurements, as reported in Figure 2b of the main text, are used to evaluate the crystal structure of 2H-$MoS_2$ flakes and QDs with respect to the bulk $MoS_2$. Bulk $MoS_2$ exhibits the characteristic XRD peaks of a hexagonal-structure polycrystalline films (JCPDS card no.77-1716).[10] The dominant (002) peak, centered at 14.4°, corresponds to the interlayer d-spacing of 0.614 nm.[11,12] In addition, various weak diffraction reflections are also observed at higher angles, *e.g.*, the ones attributed to the (100), (103), (006), (105), and (008) planes,[Error! Bookmark not defined.,12] which are characteristic of polycrystalline $MoS_2$.[12] For the 2H-$MoS_2$ flakes, the intensity of the (002) peak increases with respect to that of bulk $MoS_2$, indicating a preferential exposure of (002) basal planes,[11] while all the other diffraction reflections almost disappear, in agreement with the single crystal structure of the few-layered 2H-$MoS_2$ flakes.[Error! Bookmark not defined.] In the case of QDs, the intensity of (002) peak is strongly reduced with respect to the bulk material and the 2H-$MoS_2$ flakes. Similar to 2H-$MoS_2$ flakes, all the other diffraction peaks disappear.[Error! Bookmark not defined.] These results indicate that the QDs have the same crystal structure of their native flakes.

### S. 3 Optical absorption spectroscopy of 2H-$MoS_2$ flakes and 2H-$MoS_2$ QDs

Figure 2c in the main text shows the absorption spectra of the as-produced 2H-$MoS_2$ flakes and QDs. For 2H-$MoS_2$ flakes, the peaks at 660 and 600 nm are ascribed to the A and B excitonic peaks, respectively, arising from the K-point of the Brillouin zone in 2H-$MoS_2$ flakes.[13,14] Their energy difference (~0.2 eV) arises from the spin-orbit splitting of the valence band in 2H-$MoS_2$ flakes.[13,15] The distinct peaks at 450 and 395 nm are assigned to direct excitonic *C* and *D* inter-band transitions between the density of state peaks in the valence and conduction bands at the M point of the Brillouin zone.[13,16] In the case of 2H-$MoS_2$ QDs, there are no characteristic excitonic peaks, and the absorption edge is shifted towards lower wavelength with respect to 2H-$MoS_2$ flakes. This effect could be ascribed to the quantum confinement effect in QDs,[14] which increases their band gap energy with the decrease of the lateral size.[15,17]

## S.4 Photoluminescence characterization of 2H-MoS$_2$ QDs

The PL spectra of 2H-MoS$_2$ QDs dispersion in IPA, collected at different excitation wavelengths (from 280 to 500 nm) are reported in Figure S1a. The PL peaks are red-shifted with increasing excitation wavelength. This excitation-dependent PL emission is ascribed to quantum confinements[18] and edge state emission effect.[19,20] The sharp small features observed in the spectra are related to the IPA solvent, as observed in its blank PL spectrum (Figure S1b).

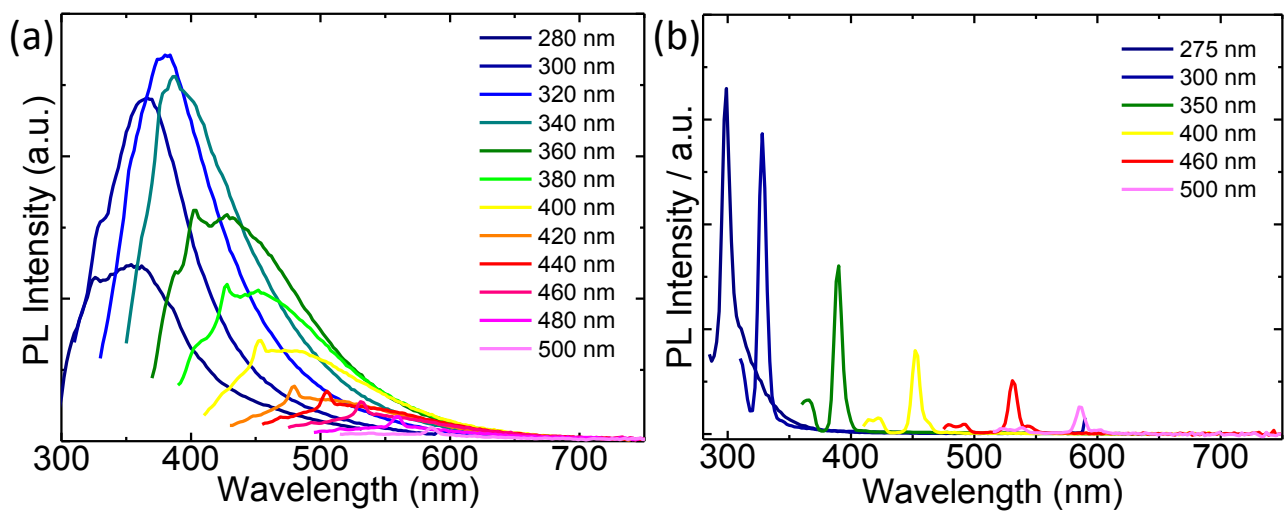

**Figure S1** (a) Photoluminescence spectra of the 2H-MoS$_2$ QDs at different excitation wavelength (*i.e.*, ranging from 280 to 500 nm). (b) Blank (control) PL measurements of IPA at different excitation wavelength.

## S.5 Raman analysis of bulk MoS$_2$, 2H-MoS$_2$ flakes, 2H-MoS$_2$ QDs and 1T-MoS$_2$ flakes

Raman spectroscopy measurements, reported in Figure 2d of the main text, are carried out to investigate the vibrational modes of 2H-MoS$_2$ flakes and QDs with respect to those of bulk MoS$_2$. The full width at half maximum of the $E_{2g}^1(\Gamma)$ and $A_{1g}(\Gamma)$ (FWHM($E_{2g}^1(\Gamma)$) and FWHM($A_{1g}(\Gamma)$), respectively) of both 2H-MoS$_2$ flakes and QDs increases of ~3 cm$^{-1}$ and ~2 cm$^{-1}$, respectively, compared to the corresponding modes of bulk MoS$_2$. The increase of FWHM($A_{1g}(\Gamma)$) for 2H-MoS$_2$ flakes and QDs is attributed to the variation of interlayer force constants between the inner and outer layers.[21]

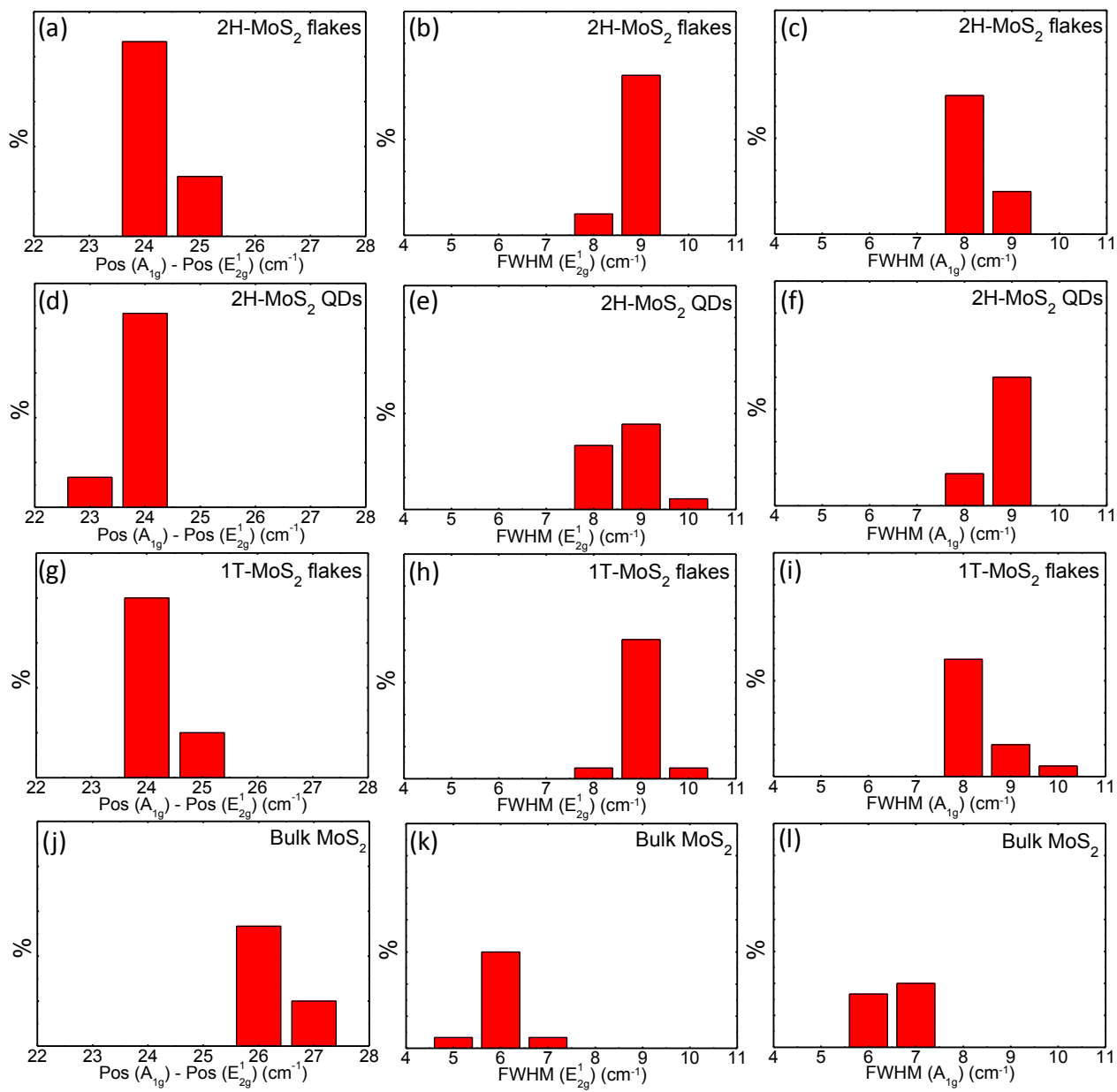

**Figure S2** Statistical Raman analysis of (a) Pos($A_{1g}$) - Pos($E_{2g}^1$), (b) FWHM($E_{2g}^1$) and (c) FWHM($A_{1g}$) for 2H-MoS$_2$ flakes; (d) Pos ($A_{1g}$) - Pos($E_{2g}^1$), e) FWHM($E_{2g}^1$) and f) FWHM($A_{1g}$) for 2H-MoS$_2$ QDs; g) Pos ($A_{1g}$) - Pos($E_{2g}^1$), h) FWHM($E_{2g}^1$) and (i) FWHM($A_{1g}$) and 1T- MoS$_2$ flakes; (j) Pos ($A_{1g}$) - Pos($E_{2g}^1$), k) FWHM($E_{2g}^1$) and (l) FWHM($A_{1g}$) for bulk MoS$_2$.

Figure S2 shows the statistical Raman analysis of the peaks position difference of the $A_{1g}(\Gamma)$ and $E_{2g}^1(\Gamma)$ modes, *i.e.* Pos($A_{1g}$) - Pos($E_{2g}^1$), FWHM($E_{2g}^1$) and FWHM($A_{1g}$) for 2H-MoS$_2$ flakes (Figure S2a, FigureS2b and FigureS2c, respectively), 2H-MoS$_2$ QDs (Figure S2d, Figure S2e and Figure S2f, respectively), 1T-MoS$_2$ flakes (Figure S2g, Figure S2h and Figure S2i, respectively) and bulk MoS$_2$ (Figure S2j, Figure S2k and Figure S2l, respectively).

Figure S3 shows the Raman spectra of bulk MoS$_2$, 2H-MoS$_2$ flakes, 2H-MoS$_2$ QDs, and 1T-MoS$_2$ flakes in the 200-900 cm$^{-1}$ range. The data do not reveal remarkable additional peaks in the 200-900 cm$^{-1}$ range related to molybdenum oxide species,[22] such as the MoO$_3$ bands located at ~285 cm$^{-1}$ ($B_{2g}$, $B_{3g}$), ~666 cm$^{-1}$ ($B_{2g}$, $B_{3g}$) and ~820 cm$^{-1}$ ($A_g$, $B_{1g}$)[23] or the MoO$_2$ band located at ~203 cm$^{-1}$, ~228 cm$^{-1}$, ~345 cm$^{-1}$, ~363 cm$^{-1}$, ~461 cm$^{-1}$, ~495 cm$^{-1}$, ~571 cm$^{-1}$, ~589 cm$^{-1}$ and ~744 cm$^{-1}$.[24] The peaks located at ~520 cm$^{-1}$ and ~303 cm$^{-1}$ are attributed to the transverse optical (TO) and the second-order transverse acoustic (2TA) phonon modes of Si[25,26,27] (samples are deposited onto Si/SiO$_2$ substrates), respectively.

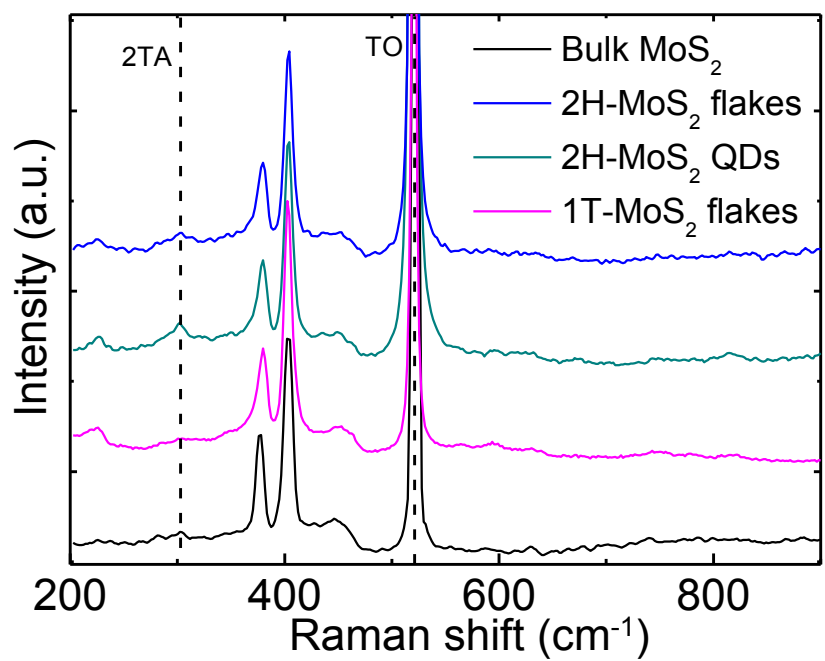

**Figure S3** Extended Raman spectra of the bulk MoS$_2$ (black), 1T-MoS$_2$ (magenta), 2H-MoS$_2$ flakes (blues) and 2H-MoS$_2$ QDs (cyan). The peaks located at ~520 cm$^{-1}$ and 303 cm$^{-1}$ are ascribed to the transverse optical (TO) and the second-order transverse acoustic (2TA) phonon modes of Si are indicated by black vertical dashed lines

## S.6 Optical, morphological and chemical characterization of graphene flakes
### S.6.1 Optical absorption spectroscopy analysis of graphene flakes

The OAS measurement of the as-produced graphene flakes dispersion in NMP (1:10 diluted) is reported in Figure S4. The peak at ~265 nm, is a signature of the van Hove singularity in the graphene density of states.[28] The concentration of graphene flakes in dispersion is determined from the optical absorption coefficient at 660 nm, using A = αlc where l [m] is the light path length, c [g L$^{-1}$] is the concentration of dispersed graphitic material, and α [L g$^{-1}$ m$^{-1}$] is the absorption coefficient, with α ~1390 L g$^{-1}$ m$^{-1}$ at 660 nm.[28,29] The obtained concentrations for the 1:10 diluted as-produced graphene flakes dispersion is 0.32 g L$^{-1}$.

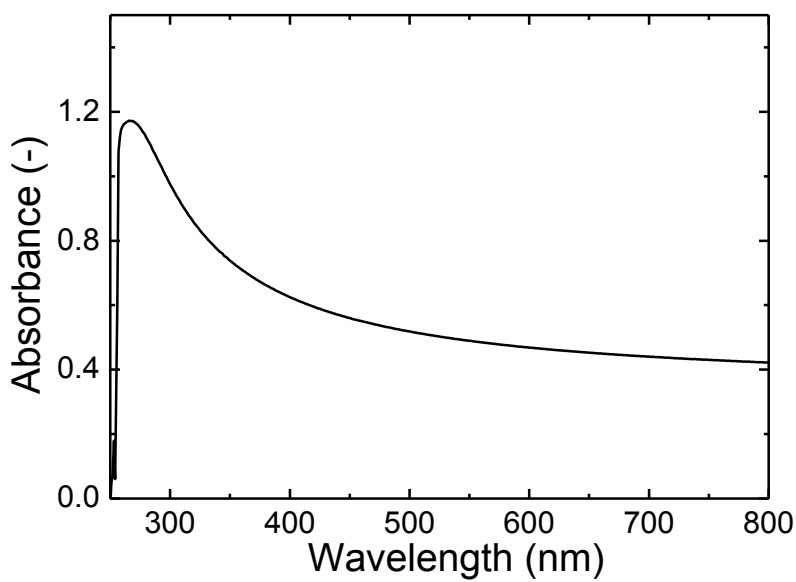

**Figure S4** Absorption spectrum of the 1:10 diluted as-produced graphene flakes dispersion in NMP.

### S.6.2. Raman spectrum and statistical Raman analysis of graphene flakes

The as-produced graphene flakes are characterized by means of Raman spectroscopy. A typical Raman spectrum of defect-free graphene shows, as fingerprints, G and D peaks.[30] The G peak corresponds to the $E_{2g}$ phonon at the Brillouin zone center.[31] The D peak is due to the breathing modes of $sp^2$ rings and requires a defect for its activation by double resonance.[30,32,33] The 2D peak is the second order of the D peak,[34] being a single peak in monolayer graphene, whereas it splits in four in bi-layer graphene, reflecting the evolution of the band structure.[30] The 2D peak is always seen, even in the absence of D peak, since no defects are required for the activation of two phonons with the same momentum, one backscattered from the other.[34] Double resonance can also happen as intra-valley process, *i.e.,* connecting two points belonging to the same cone around K or K'.[34] This process gives rise to the D' peak for defective graphene.[34] The D+D' is the combination mode of D and D' while the 2D' is the second order of the D'.[34] As in the case of 2D, 2D' is always seen even when the D' peak is not present.[34] Figure S5a reports a representative Raman spectrum of the as-produced graphene flakes, showing all the bands above described.

The statistical analysis of the position of G (Pos(G)) (Figure S5b), the full width half maximum of G (FWHM(G)) (Figure S5c), the position of 2D (Pos(2D)) (Figure S5d), the full width half maximum of 2D (FWHM(2D)) (Figure S5e), the intensity ratio between the 2D and G peaks (I(2D)/I(G)) (Figure S5f) and the intensity ratio between the D and G peaks (I(D)/(IG)) (Figure S5i) gives useful quantitative information on the graphene flakes characteristics. In particular, the Pos(2D) is at ~2700 cm$^{-1}$ (Figure S5d) while the FWHM(2D) ranges from 60 to 75 cm$^{-1}$ (Figure S5e). These values are ascribed to few-layer graphene (FLG).[4,30,35] The I(2D)/I(G) varies from 0.6 to 1.2 (Figure S5f), as expected from a combination of single-layer graphene (SLG) and FLG.[30,36] The presence of D and D' indicate, as discussed for D, the defective nature of the graphene flakes.[34,37,38,39] Previous studies on graphene flakes produced by LPE have shown that these defects are predominantly located at the edges, while the basal plane of the flakes is defect-free.[38,39,] This is demonstrated by the absence of correlation between I(D)/I(G) and FWHM(G).[37,39]

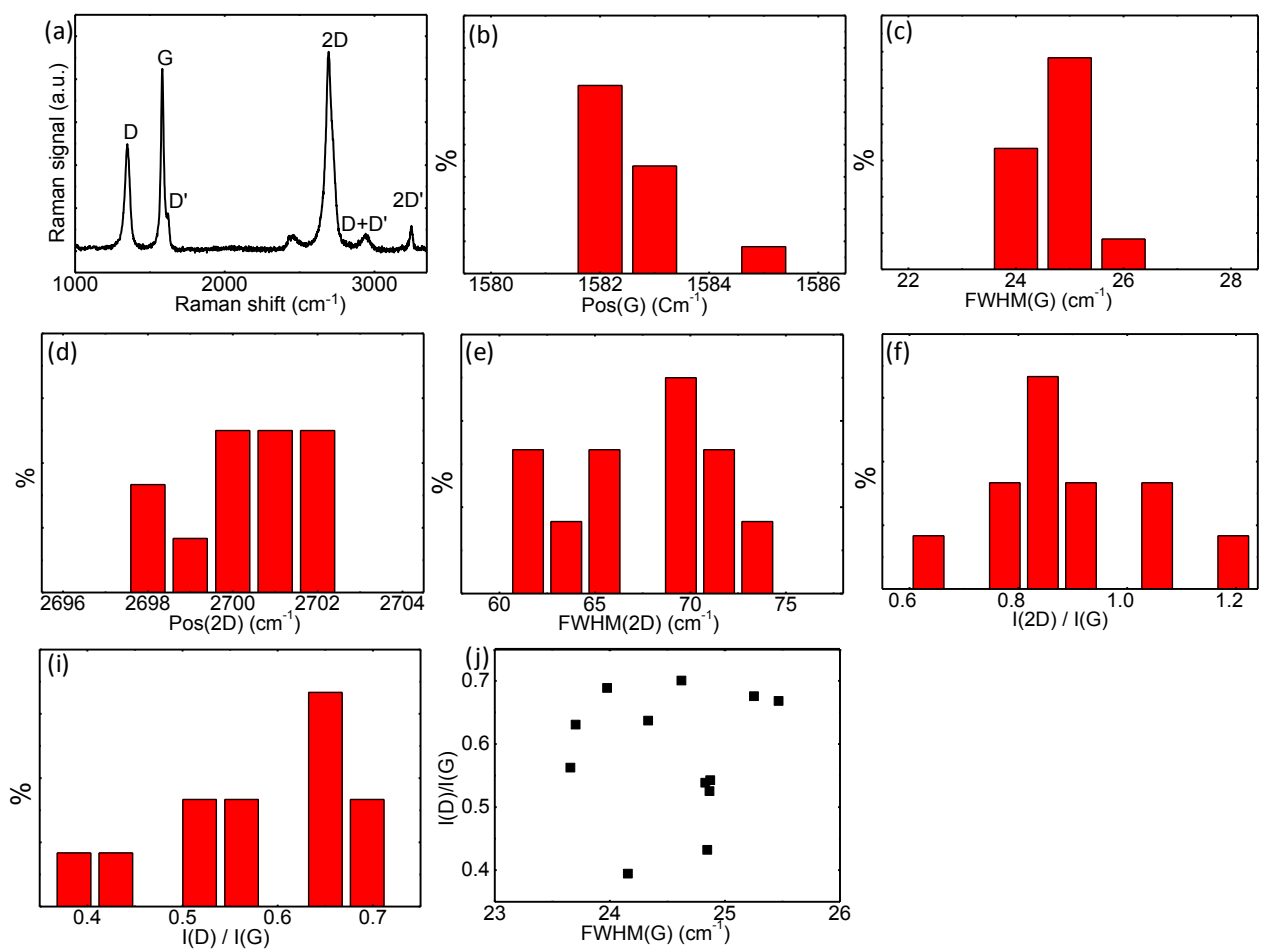

**Figure S5** (a) Representative Raman spectrum of the as produced SLG/FLG by LPE in NMP. The D, G, D', 2D, D+D' and 2D' bands are also denoted. (b) Statistical Raman analysis of the Pos (G), (c) FWHM(G), (d) Pos(2D), (e) FWHM(2D), (f) I(2D)/I(G), (i) I(D)/I(G) and (j) I(D)/I(G) *vs.* FWHM(G) plot.

Figure S5i shows the statistical analysis of I(D)/I(G), which varies between 0.3 and 0.7, while Figure S5j does not show, in agreement with literature data,[37,39] any correlation between I(D)/I(G) and FWHM(G), thus proving defect-free basal planes of the as produced graphene flakes.

### S.6.3 Morphological characterization of graphene flakes

The morphology of the as-produced graphene flakes is characterized by means of transmission electron microscopy (TEM) and AFM. Figure S6a shows a representative TEM image of graphene flakes, which have irregular shape and rippled morphology. Statistical TEM analysis of the flakes lateral dimension indicates values distributed in the range of 200-1500 nm and an average value of ~450 nm.

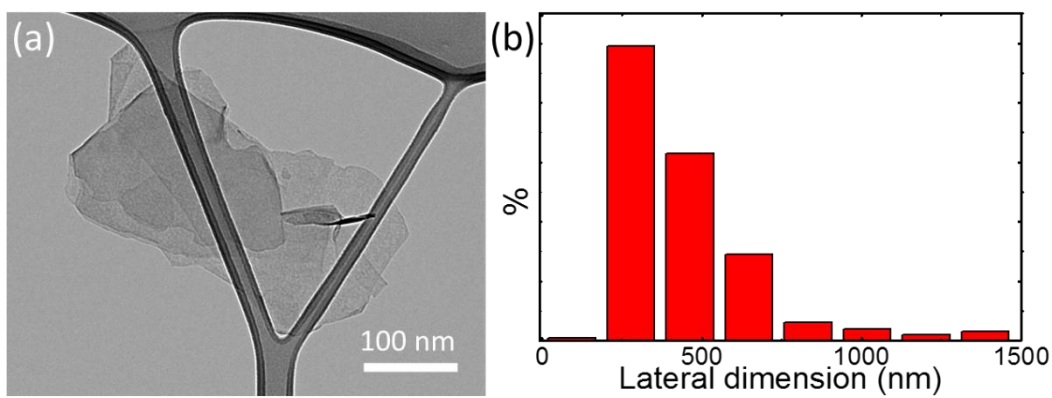

**Figure S6** (a) TEM images of the as-produced graphene flakes and (b) TEM statistical analysis of their lateral dimension.

Figure S7a shows a representative AFM image of graphene flakes. The main thickness distribution is in the 0.5-4 nm range (Figure S7b), with the presence of few thicker flakes (>5 nm). Thus, the sample is mostly composed by a combination of SLG and FLG flakes, in agreement with Raman spectroscopy analysis (see Section S6.2).

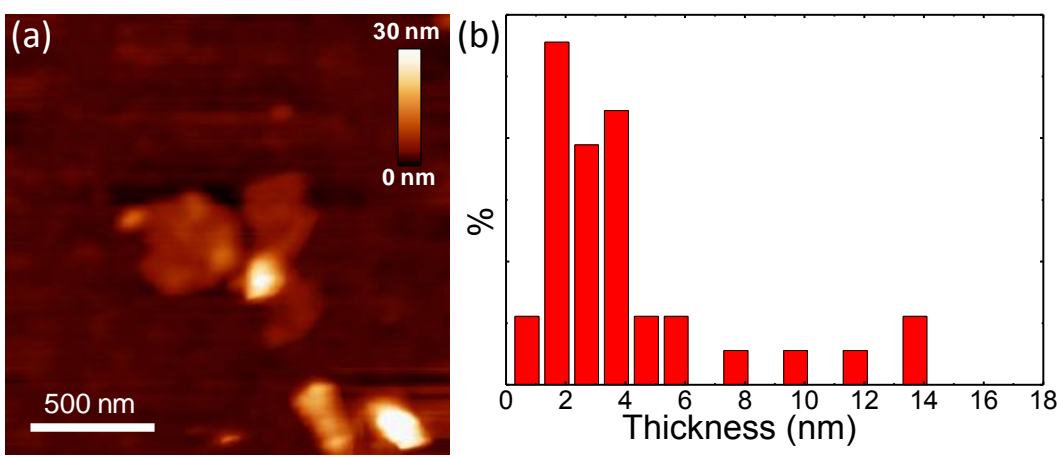

**Figure S7** (a) AFM images of the as-produced graphene flakes and (b) AFM statistical analysis of their lateral dimension.

## S.6.4 X-ray photoemission spectroscopy analysis of graphene flakes

X-ray photoelectron spectroscopy (XPS) measurements are carried out on as-produced graphene flakes to ascertain their chemical composition. The results are shown in Figure S8. The C 1s spectrum of the graphene flakes (Figure S8a), shows the presence of oxidized C-O and C=O groups at binding energies 286.4 eV and 288.3 eV respectively.[40] The percentage content (%c) of C=O and C-O is ~8%. However, these groups are also attributed to the presence of residual solvent molecule of NMP. In fact, N 1s spectrum (Figure S8b) indicates a %c of NMP ~3.5%. Take into account the NMP contribution in the %c of the oxidized groups, these results confirm that high-quality graphene flakes (%c >95%) are effectively obtained by LPE in NMP, in acgreement with previous studies.[41]

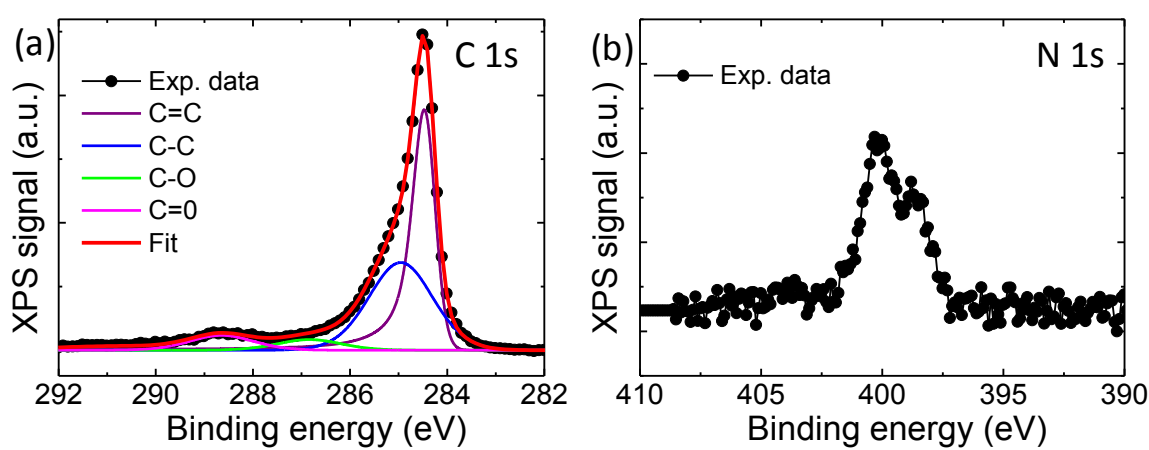

**Figure S8** (a) C 1s and b) N 1s XPS spectra of the graphene flakes sample produced by LPE of graphite in NMP. The deconvolution of C 1s XPS spectra is also shown, indicating the contribution of C=C (purple), C-C (blue), C-O (green), C=O (magenta).

## S.7 Morphology characterization of the hybrid graphene flakes/2H-MoS$_2$ flakes and graphene flakes/2H-MoS$_2$ QDs heterostructures

The morphology of the different flexible heterostructures, fabricated by the sequential deposition of graphene and MoS$_2$ dispersions on nylon membranes, is analysed by using AFM. Figure S9 shows the representative 1.5 × 1.5 µm$^2$ AFM topographies of the graphene, graphene/2H-MoS$_2$ flakes, and graphene/2H-MoS$_2$ QDs flexible electrodes (Figure S9a,d,g), as well as, their respective AFM phase images (Figure S9b,e,h) and AFM 3D images (Figure S9c,f,i). The analysis of the roughness derived from Figure S9 a, d and g reveals the lowest roughness (Ra = 10 nm; RMS = 15nm) in the case of graphene/2H-MoS$_2$ QDs where the presence of graphene flakes is not observed. In the other two cases, the uniform coverage of the surface with layered material reported roughness of Ra = 16 nm; RMS = 20nm and Ra = 25 nm; RMS = 31 nm for graphene and graphene/2H-MoS$_2$ flakes electrodes, respectively. The homogeneous coverages of the layered material, for graphene and graphene/2H-MoS$_2$ flakes, and QDs or QDs aggregates in the case of graphene/2H-MoS$_2$ QDs heterostructures is also confirmed by the AFM phase images displayed in Figure S9b, Figure S9e and Figure S9h, respectively. In fact, these images show the domains of the different overlay materials of the electrodes over the entire imaged areas (1.5 × 1.5 µm$^2$).

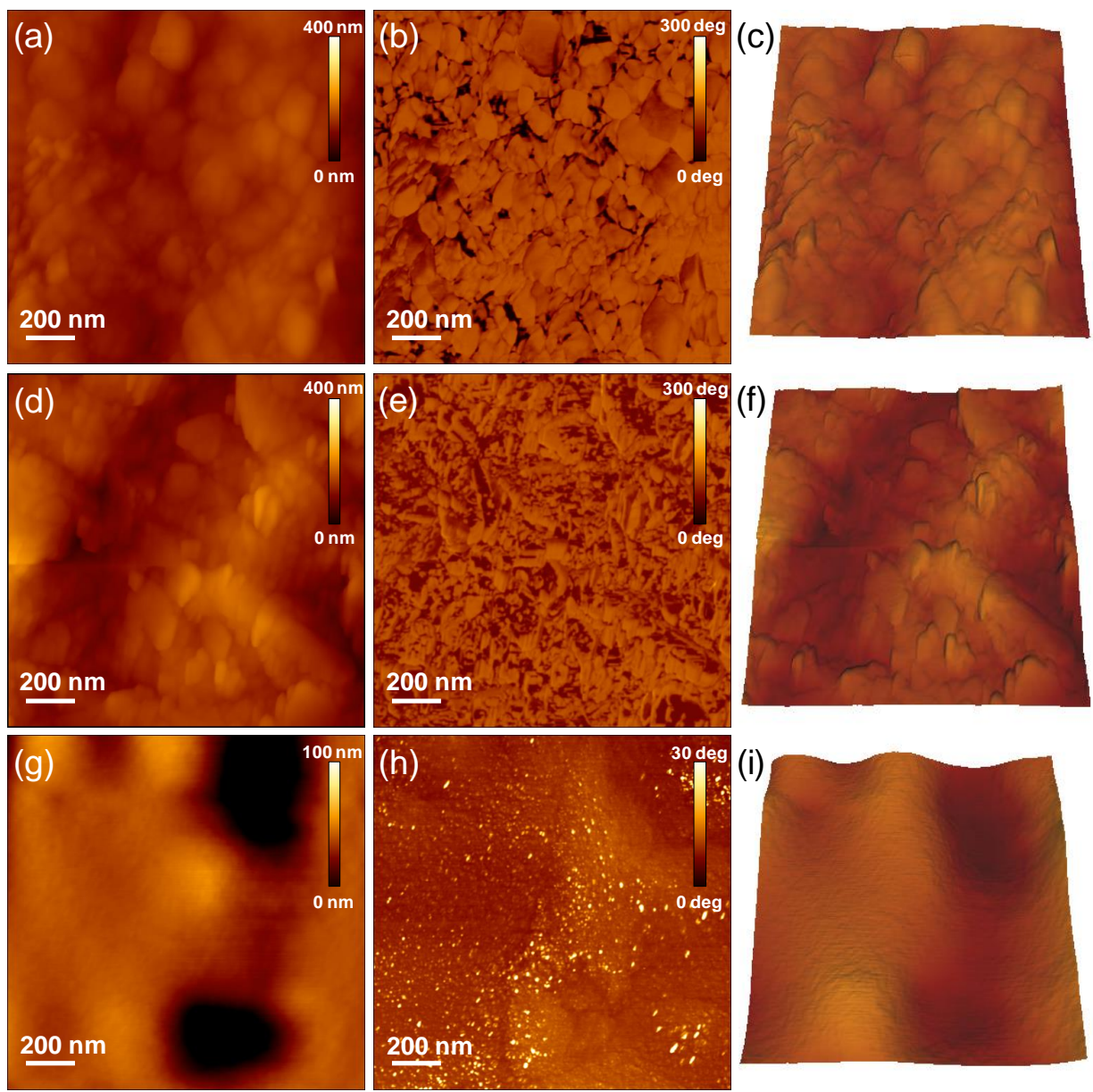

**Figure S9** Atomic force microscopy characterization of the (a-c) graphene, (d-f) graphene/2H-MoS$_2$ flakes, and (g-i) graphene/2H-MoS$_2$ QDs flexible electrodes. The AFM topography images of (a) graphene, (d) graphene/2H-MoS$_2$ flakes, and (g) graphene/2H-MoS$_2$ QDs flexible electrodes show layered material in the first two cases, while in the case of graphene/2H-MoS$_2$ QDs it is observed a smoother surface covered by the presence of QDs. The AFM phase images of graphene,

graphene/2H-MoS$_2$ flakes and graphene/2H-MoS$_2$ QDs flexible electrodes are presented in (b), (e) and (h) respectively, while the corresponding 3D images are shown in (c), (f) and (i), respectively.

## S.8 X-ray photoemission spectroscopy analysis of 1T-MoS$_2$ flakes and MoS$_2$ QDs produced from 1T-MoS$_2$ flakes

The 1T-MoS$_2$ flakes are obtained by chemical lithium intercalation method.[Error! Bookmark not defined.] This method results in loss of pristine semiconducting properties of 2H-MoS$_2$ flakes due to structural changes that occur during Li intercalation, *i.e.,* 1T-MoS$_2$ flakes formation.[42,43,44,45,46] The morphological and optical characterization of the as-produced 1T-MoS$_2$ flakes has been recently reported by our group.[47] As main results, the TEM analysis of the 1T-MoS$_2$ flakes indicated lateral size in the 30-800 nm range (average value ~275 nm), while AFM analysis revealed average thickness of 2.3 ± 1.6 nm. Moreover, XPS analysis revealed that the resulting MoS$_2$ flakes are a mixture of both 2H and 1T phase. However, the metastable metallic 1T phase dominate the electrocatalytic properties of the as-exfoliated material,[42-46] but mild annealing (~100 °C) leads to gradual restoration of the semiconducting phase.[42] Figure S10a reports the XPS spectra of the as produced 1T-MoS$_2$ flakes. The peaks located at ~229 eV and at ~232 eV are assigned to Mo 3d of MoS$_2$ and fitted by two components, which are attributed to the 2H (green line) and 1T phase (cyan line) of MoS$_2$ flakes, respectively. Figure S10b shows the XPS spectra of MoS$_2$ QDs produced starting from the 1T-MoS$_2$ flakes. Clearly, the 1T phase contribution is reduced with respect to the one observed in 1T-MoS$_2$ flakes, indicating that the solvothermal treatment causes a 1T-to-2H phase conversion. These results prove that it is challenging to produce 1T-MoS$_2$ QDs from 1T-MoS$_2$ flakes because of the intrinsic metastable nature of the latter.[48,49,50]

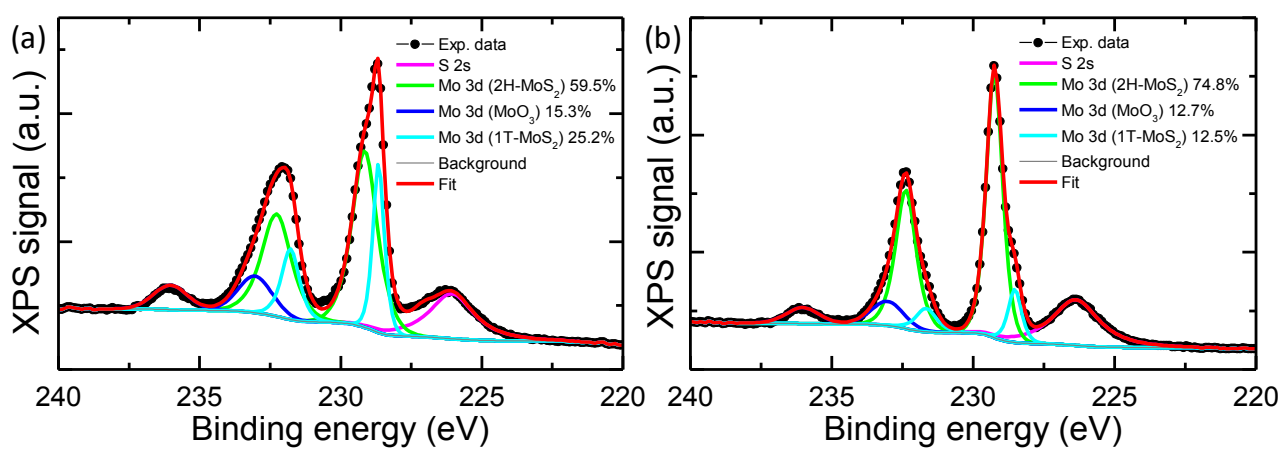

**Figure S10** Mo 3d and S 2s XPS spectra for (a) 1T-MoS$_2$ flakes and (b) the MoS$_2$ QDs derived from 1T-MoS$_2$ flakes. Their deconvolutions are shown, indicating the contribution of both 1T and 2H phase of MoS$_2$ for Mo 3d peaks (solid cyan and green lines, respectively). The S 2s band of MoS$_2$ and Mo 3d bands of MoO$_3$ are also evidenced (solid magenta and blue lines, respectively). The percentage

contents of Mo 3d bands attributed to 2H-MoS$_2$, 1T-MoS$_2$ and MoO$_3$ are also reported in the figure legends.

## S.9 Stability tests of graphene/2H-MoS$_2$ flakes and graphene/2H-MoS$_2$ QDs in HER-conditions

The stability of the graphene/2H-MoS$_2$ flakes and graphene/2H-MoS$_2$ QDs in hydrogen evolution reaction (HER)-condition is evaluated by chronoamperometry measurements (*j-t* curves) at -0.5 V *vs.* RHE. Figure S11 shows the results of the tests over 200 min of continuous operation. The graphene/2H-MoS$_2$ QDs show a progressive HER activation, with an increase of the current of ~10% after 200 min, while the current of graphene/2H-MoS$_2$ flakes decrease of ~4% with respect to the starting values. These results suggest that the catalytic edge sites of 2H-MoS$_2$ QDs are more resistant toward oxidative/degradation processes, which passivate the HER catalytic sites, with respect to those of 2H-MoS$_2$ flakes. In fact, density functional theory calculation have shown that oxidation energies for MoS$_2$ flakes depend on the local competition of binding energy of the covalent bonds at the edge sites,[51] whose nature can be different for 2D and 0D nanostructures.[52]

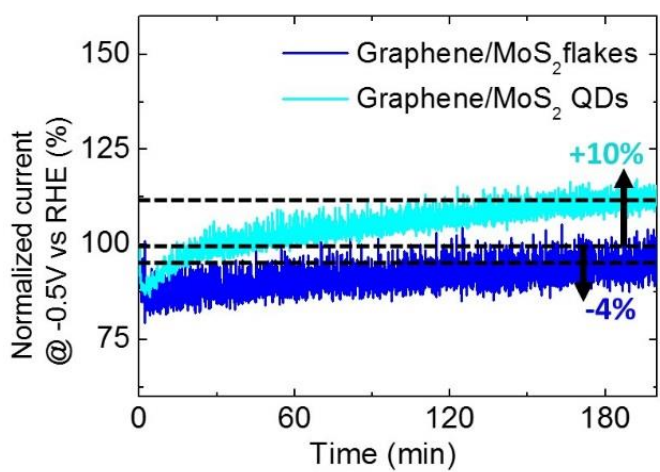

**Figure S11** Chronoamperometry measurements (j-t curves) at -0.5 V *vs.* RHE of the graphene/2H-MoS$_2$ flakes (blue lines) and graphene/2H-MoS$_2$ QDs (cyan line), over 200 min.

## S.10 Comparison of the HER electrocatalytic activity of the as-produced devices in the literature context

As pointed out in the conclusion of the main text, the HER electrocatalytic performance of our systems (*e.g.* GC/2H-MoS$_2$ QDs and the graphene/2H-MoS$_2$ QDs approach that of several MoS$_2$-based catalyst reported in literature, overcoming those of recent MoS$_2$ flakes or MoS$_2$ QDs synthetized by scalable routes compatible with high-throughput industrial processes.[14,18,53,54,55] The comparison of the HER electrocatalytic activity of our system with other relevant reported findings (not strictly referring to material synthesis compatible with high-throughput industrial processes) is reported in Table S1.

**Table S1** Comparison of HER electrocatalytic activity of our systems with other relevant reported findings

| HER electrocatalyst | $\eta_{10}$ (V) | Tafel slope (mV/dec) | reference |
|---|---|---|---|
| Graphene (GC)/2H-MoS$_2$ QDs | 0.136 (0.312) | 141 (78) | this work |
| MoS$_2$ nanosheets | ~0.43 | 115 | 53 |
| MoS$_2$ dots on Au | 0.16 | 82 | 12 |
| MoS$_2$ dots/nanosheet hybrid on Au | 0.19 | 74 | 18 |
| MoS$_2$ nanodots | ~0.28 | 61 | 14 |
| Thermal texturized MoS$_2$ | 0.17 | ~60-70 | 54 |
| Microdomain reaction-produced MoS$_2$ nanosheets | ~0.19 | 68 | 55 |
| H$_2$-annealed MoS$_2$ monolayer | <0.55 | 147 | 56 |
| 1T-MoS$_2$ nanosheets | ~0.20 | 40 | 42 |
| Double-gyroid MoS$_2$ films | <-0.20 | 50 | 57 |
| Edge-exposed MoS$_2$ nano-assembled structures | ~0-18/0.19 | 100 | 58 |
| stepped edge surface terminated MoS$_2$ sheet arrays | 0.10 | 59 | 59 |
| flat edge surface terminated MoS$_2$ sheet arrays | 0.14 | 69 | 59 |
| MoS$_2$/RGO hybrid | ~0.15 | 41 | 60 |

## Acknowledgments
We thank the Electron Microscopy facility - Istituto Italiano di Tecnologia, for support in TEM data acquisition.

## References
(1) Nicolosi V.; Chhowalla M.; Kanatzidis M. G.; Strano M. S.; Coleman J. N. Liquid exfoliation of layered materials. *Science*, **2013**, *340*, 1226419.

(2) Bonaccorso, F.; Lombardo, A.; Hasan, T.; Sun, Z.; Colombo, L.; Ferrari, A. C. Production and processing of graphene and 2d crystals. *Mater. Today*, **2012**, *15*, 564-589.

(3) Bonaccorso, F.; Bartolotta, A.; Coleman, J. N.; Backes, C. 2D-Crystal-Based Functional Inks. *Adv. Mater.*, **2016**, *28*, 6136.

(4) Maragó, O. M.; Bonaccorso, F.; Saija, R.; Privitera, G.; Gucciardi, P. G.; Iati, M. A.; Calogero, G.; Jones, P. H.; Borghese, F.; Denti, P.; Nicolosi, V. Brownian motion of graphene. *ACS Nano*, **2010**, *4*, 7515-7523.

(5) Casaluci, S.; Gemmi, M.; Pellegrini, V.; Di Carlo, A.; Bonaccorso, F. Graphene-based large area dye-sensitized solar cell modules. *Nanoscale*, **2016**, *8*, 5368.

(6) Yuwen L.; Yu H.; Yang X.; Zhou J.; Zhang Q.; Zhang Y.; Wang L. Rapid preparation of single-layer transition metal dichalcogenide nanosheets via ultrasonication enhanced lithium intercalation. *Chem. Comm.*, **2016**, *52*, 529-532.

(7) Jung Y.; Zhoub Y.; Cha J. J. Intercalation in two-dimensional transition metal chalcogenides. *Inorg. Chem. Front.*, **2016**, *3*, 452-463.

(8) Vilekar S. A.; Fishtik I.; Datta R. Kinetics of the hydrogen electrode reaction. *J. Electrochem. Soc.*, **2010**, *157*, B1040-B1050.

(9) Conway B. E.; Tilak B. V. Interfacial processes involving electrocatalytic evolution and oxidation of $H_2$, and the role of chemisorbed H. *Electrochim. Acta*, **2002**, *47*, 3571-3594.

(10) Joensen P.; Crozier E. D.; Alberding N.; Frindt R. F. A study of single-layer and restacked $MoS_2$ by X-ray diffraction and X-ray absorption spectroscopy, *J. Phys. C: Solid State Phys.*, **1987**, *20*, 4043-4053

(11) Yue N.; Weicheng J.; Rongguo W.; Guomin D.; Yifan H. Hybrid nanostructures combining graphene–$MoS_2$ quantum dots for gas sensing. *J. Mater. Chem. A*, **2016**, *4*, 8198-8203

(12) Wang T.; Gao D.; Zhu J.; Papakonstantinou P.; Li Y.; Li M. X. Size-Dependent Enhancement of Electrocatalytic Oxygen-Reduction and Hydrogen-Evolution Performance of $MoS_2$ Particles. *Chem.-Eur. J.*, **2013**, *19*, 11939-11948

(13) Mak K. F.; Lee C.; Hone J.; Shan J.; Heinz T. F. Atomically thin $MoS_2$: a new direct-gap semiconductor. *Phys. Rev. Lett.*, **2010**, *105*, 136805.

(14) Benson, J.; Li, M.; Wang, S.; Wang, P.; Papakonstantinou, P. Electrocatalytic hydrogen evolution reaction on edges of a few layer molybdenum disulfide nanodots, *ACS Appl. Mater. Interfaces*, **2015**, *7*,14113-14122.

(15) Wilcoxon J. P.; Samara G. A. Strong quantum-size effects in a layered semiconductor: MoS2 nanoclusters. *Phys. Rev. B: Condens. Matter Mater. Phys.*, **1995**, *51*, 7299.

(16) Wilcoxon J. P.; Newcomer P. P.; Samara G. A.; Synthesis and optical properties of MoS2 and isomorphous nanoclusters in the quantum confinement regime. *J. Appl. Phys.*, **1997**, *81*, 7934-7944

(17) Muscuso L.; Cravanzola S.; Cesano F.; Scarano D.; Zecchina A. Optical, Vibrational, and structural properties of MoS2 nanoparticles obtained by exfoliation and fragmentation via ultrasound cavitation in isopropyl alcohol., *J. Phys. Chem. C*, **2015**, *119*, 3791-3801.

(18) Gopalakrishnan D.; Damien D.; Shaijumon M. M. MoS2 quantum dot-interspersed exfoliated MoS2 nanosheets. *ACS Nano*, **2014**, *8*, 5297-5303.

(19) Gopalakrishnan D.; Damien D.; Li B.; Gullappalli H.; Pillai V. K.; Ajayan P. M.; Shaijumon M. M. Electrochemical synthesis of luminescent MoS2 quantum dots. *Chem. Commun.*, **2015**, *51*, 6293-6296

(20) Gan Z.; Xu H.; Hao Y. Mechanism for excitation-dependent photoluminescence from graphene quantum dots and other graphene oxide derivates: consensus, debates and challenges. *Nanoscale*, **2016**, *8*, 7794-7807

(21) Lee C.; Yan H.; Brus L. E.; Heinz T. F.; Hone J.; Ryu S. Anomalous lattice vibrations of single-and few-layer MoS2. *ACS Nano*, **2010**, *4*, 2695-2700.

22) Dieterle M.; Mestl G. Raman spectroscopy of molybdenum oxides Part II. Resonance Raman spectroscopic characterization of the molybdenum oxides Mo4O11 and MoO2. *Phys. Chem. Chem. Phys.*, **2002**, *4*, 822-826.

(23) Py M. A.; Maschke K. Intra-and interlayer contributions to the lattice vibrations in MoO3. *Phys. B*, **1981**, *105*, 370-374.

(24) Spevack P. A.; McIntyre N.S. Thermal reduction of molybdenum trioxide. *J. Phys. Chem.*, **1992**, *96*, 9029-9035.

(25) Zhang S. L.; Wang X.; Ho K.; Li J.; Diao P.; Cai S. Raman spectra in a broad frequency region of p-type porous silicon. *J. Appl. Phys.*, **1994**, *76*, 3016-3019

(26) Richter H.; Wang Z. P.; Ley L. The one phonon Raman spectrum in microcrystalline silicon. *Solid State Commun.*, **1981**, *39*, 625-629

(27) Temple P. A.; Hathaway C. E. Multiphonon Raman spectrum of silicon. *Phis. Rev. B*, **1973**, *7*, 3685.

(28) Kravets V. G.; Grigorenko A. N.; Nair R. R.; Blake P.; Anissimova S.; Novoselov K. S.; Geim A. K. Spectroscopic ellipsometry of graphene and an exciton-shifted van Hove peak in absorption. *Phys. Rev. B*, **2010**, *81* 155413.

(29) Lotya M.; Hernandez Y.; King P. J.; Smith R. J.; Nicolosi V.; Karlsson L. S.; Blighe F. M.; De S.; Zhiming W.; McGovern I. T.; Duesberg G. S.; Coleman J. N. Liquid phase production of graphene by exfoliation of graphite in surfactant/water solutions. *J. Am. Chem. Soc.*, **2009**, *131,* 3611-3620.

(30) Ferrari A. C.; Meyer J. C.; Scardaci V.; Casiraghi C.; Lazzeri M.; Mauri F.; Piscanec S.; Jiang D.; Novoselov K. S.; Roth S.; Geim A. K. Raman spectrum of graphene and graphene layers. *Phys. Rev. Lett.*, **2006**, *97*, 187401.

(31) Ferrari A. C.; Basko D. M. Raman spectroscopy as a versatile tool for studying the properties of graphene. *Nat. Nanotechnol.*, **2013**, *8*, 235-246

(32 Ferrari A. C.; Robertson J. Interpretation of Raman spectra of disordered and amorphous carbon. *Phys. Rev. B*, **2000**, *61*, 14095-14107

(33) Ferrari A. C.; Robertson J. Resonant Raman spectroscopy of disordered, amorphous, and diamondlike carbon. *Phys. Rev. B*, **2001**, *64*, 075414.

(34) Su C.-Y.; Xu Y.; Zhang W.; Zhao J.; Tang X.; Tsai C.-H.; Li L.-J. Electrical and spectroscopic characterizations of ultra-large reduced graphene oxide monolayers. *Chem. Mater.*, **2009**, *21*, 5674.

(35) Hassoun J.; Bonaccorso F.; Agostini M.; Angelucci M.; Betti M. G.; Cingolani R.; Gemmi M.; Mariani C.; Panero S.; Pellegrini V.; Scrosati B. An advanced lithium-ion battery based on a graphene anode and a lithium iron phosphate cathode. *Nano Lett.*, **2014**, *14*, 4901-4906

(36) Graf D.; Molitor F.; Ensslin K.; Stampfer C.; Jungen A.; Hierold C.; Wirtz L. Spatially resolved Raman spectroscopy of single-and few-layer graphene. *Nano Lett.*, **2007**, *7*, 238-242

(37) Bracamonte M. V.; Lacconi G. I.; Urreta S. E.; Foa Torres L. E. F. On the nature of defects in liquid-phase exfoliated graphene. *J. Phys. Chem. C*, **2014**, *118*, 15455-15495.

(38) Lotya M.; Hernandez Y.; King P. J.; Smith R. J.; Nicolosi V.; Karlsson L. S.; Blighe F. M.; De S.; Wang Z. M.; McGovern I. T.; Duesberg G. S.; Coleman J. N. *J. Am. Chem. Soc.*, **2009**, *131*, 3611-3620.

(39) Coleman J. N. Liquid phase production of graphene by exfoliation of graphite in surfactant/water solutions. *Acc. Chem. Res.*, **2013**, *46*, 14.

(40) Yang D.; Velamakanni A.; Bozoklu G.; Park S.; Stoller M.; Piner R. D.; Stankovich S.; Jung I.; Field D.; Ventrice C. A.; Ruoff R. S. Chemical analysis of graphene oxide films after heat and chemical treatments by X-ray photoelectron and Micro-Raman spectroscopy. *Carbon*, **2009**, *47*, 145.

(41) Buzio, R.; Gerbi, A.; Bernini, C.; Del Rio Castillo, A. E..; Palazon, F.; Siri, A. S.; Pellegrini, V.; Pellegrino, L.; Bonaccorso, F. Ultralow friction of ink-jet printed graphene flakes. *Nanoscale*, **2017**, *in press*.

(42) Voiry D.; Salehi M.; Silva R.; Fujita T.; Chen M.; Asefa T.; Shenoy V. B.; Eda G.; Chhowalla M. Conducting MoS2 nanosheets as catalysts for hydrogen evolution reaction. *Nano Lett.*, **2013**, *13*, 6222-6227.

(43) Ambrosi A.; Sofer Z.; Pumera M. Lithium intercalation compound dramatically influences the electrochemical properties of exfoliated MoS2. *Small*, **2015**, *11*, 5, 605-612.

(44) Py M. A.; Haering R. R. Structural destabilization induced by lithium intercalation in MoS2 and related compounds. *Can. J. Phys.*, **1983**, *61*, 76-84

(45) Kertesz M.; Hoffmann R. Octahedral vs. trigonal-prismatic coordination and clustering in transition-metal dichalcogenides. *J. Am. Chem. Soc.*, **1984**, *106*, 3453-3460.

(46) Chhowalla M.; Shin H. S.; Eda G.; Li L.-J.; Loh K. P.; Zhang H. The chemistry of two-dimensional layered transition metal dichalcogenide nanosheets. *Nature Chem.*, **2013**, *5*, 263.

(47) Bellani S.; Najafi L.; Capasso A.; Del Rio Castillo A. E.; Antognazza M. R.; Bonaccorso F. Few-layer MoS2 flakes as a hole-selective layer for solution-processed hybrid organic hydrogen-evolving photocathodes. *J. Mater. Chem. A*, **2017**, 5, 4384-4396.

(48) Gao G.; Jiao Y.; Ma F.; Jiao Y.; Waclawik E.; Du A. Charge mediated semiconducting-to-metallic phase transition in molybdenum disulfide monolayer and hydrogen evolution reaction in new 1T' phase. *J. Phys. Chem. C*, **2015**, *119*, 13124-13128.

(49) Voiry D.; Yamaguchi H.; Li J.; Silva R.; Alves D. C. B.; Fujita T.; Chen M.; Asefa T.; Shenoy V. B.; Eda G.; Chhowalla M. Enhanced catalytic activity in strained chemically exfoliated WS2 nanosheets for hydrogen evolution. *Nature Mater.,* **2013**, *12*, 850-855.

(50) Tsai H. L.; Heising J.; Schindler J. L.; Kannewurf C. R.; Kanatzidis M. G. Exfoliated-restacked phase of WS2. *Chem. Mater.*, **1997**, *9*, 879-882.

(51) Liang T.; Sawyer W. G.; Perry S. S.; Sinnott S. B.; Phillpot S. R. Energetics of Oxidation in MoS2 Nanoparticles by Density Functional Theory. *J. Phys. Chem. C*, **2011**, *115*, 10606-10616.

(52) Mukherjee S.; Maiti R.; Katiyar A. K.; Das S.; Ray S. K. Novel Colloidal MoS2 Quantum Dot Heterojunctions on Silicon Platforms for Multifunctional Optoelectronic Devices. *Sci. Rep.*, **2016**, *6*, 29016.

(53) Varrla E.; Backes C.; Paton K. R.; Harvey A.; Gholamvand Z.; McCauley J.; Coleman J. N.; Large-scale production of size-controlled MoS2 nanosheets by shear exfoliation. *Chem. Mater.*, **2015**, *27*, 1129-1139

(54) Kiriya D.; Lobaccaro P.; Nyein H. Y. Y.; Taheri P.; Hettick M.; Shiraki H.; Sutter-Fella C. M.; Zhao P.; Gao W.; Maboudian R.; Ager J. W.; Javey A. General Thermal Texturization Process of MoS2 for Efficient Electrocatalytic Hydrogen Evolution Reaction. *Nano Lett.*, **2016**, *16*, 4047-4053.

(55) Wu Z.; Fang B.; Wang Z.; Wang C.; Liu Z.; Liu F.; Wang W.; Alfantazi A.; Wang D.; Wilkinson D. P. MoS2 nanosheets: a designed structure with high active site density for the hydrogen evolution reaction. *ACS Catal.*, **2013**, *3*, 2101-2107.

(56) Ye G.; Gong Y.; Lin J.; Li B.; He Y.; Pantelides S. T.; Zhou W.; Vajtai R.; Ajayan P. M. Defects engineered monolayer MoS2 for improved hydrogen evolution reaction. *Nano Lett*., **2016**, *16*, 1097-1103.

(57) Kibsgaard J.; Chen Z.; Reinecke B. N.; Jaramillo T. F. Engineering the surface structure of MoS2 to preferentially expose active edge sites for electrocatalysis. *Nat. Mater.*, **2012**, *11*, 963-969.

(58) Chung D. Y.; Park S. K.; Chung Y. H.; Yu S. H.; Lim D. H.; Jung N.; Ham H. C.; Park H. Y.; Piao Y.; Yoo S. J.; Sung Y. E. Edge-exposed MoS2 nano-assembled structures as efficient electrocatalysts for hydrogen evolution reaction*Nanoscale*, **2014**, *6*, 2131-2136

(59) Hu J.; Huang B.; Zhang C.; Wang Z.; An Y.; Zhou D.; Lin H.; Leung M. K.; Yang S. Engineering stepped edge surface structures of MoS 2 sheet stacks to accelerate the hydrogen evolution reaction. *Energy Environ. Sci.*, **2017**, *10*, 593.

(60) Li Y.; Wang H.; Xie L.; Liang Y.; Hong G.; Dai H. MoS2 nanoparticles grown on graphene: an advanced catalyst for the hydrogen evolution reaction. *J. Am. Chem. Soc.*, **2011**, *133*, 7296-7299.